\documentclass[reprint,superscriptaddress, amsmath, amssymb,  aps, floatfix, longbibliography, a4paper]{revtex4-2}

\usepackage{amsmath}
\usepackage{graphicx}
\usepackage{multirow}
\usepackage{xcolor}
\usepackage{dcolumn}
\usepackage{bm}
\usepackage{float}
\usepackage{comment}
\usepackage[sectionbib,globalcitecopy]{bibunits}

\begin{document}

\title{Magnetic order in 2D antiferromagnets revealed by spontaneous anisotropic magnetostriction}

\affiliation{Kavli Institute of Nanoscience, Delft University of Technology, Lorentzweg 1,\\ 2628 CJ, Delft, The Netherlands}
\affiliation{Instituto de Ciencia Molecular (ICMol), Universitat de Val\`{e}ncia, c/Catedr\'{a}tico Jos\'{e} Beltr\'{a}n 2,\\ 46980 Paterna, Spain}
\affiliation{Department of Precision and Microsystems Engineering, Delft University of Technology,
	Mekelweg 2,\\ 2628 CD, Delft, The Netherlands}

\author{Maurits J. A. Houmes}
\thanks{These authors contributed equally.}
\affiliation{Kavli Institute of Nanoscience, Delft University of Technology, Lorentzweg 1,\\ 2628 CJ, Delft, The Netherlands}
\author{Gabriele Baglioni}
\thanks{These authors contributed equally.}
\affiliation{Kavli Institute of Nanoscience, Delft University of Technology, Lorentzweg 1,\\ 2628 CJ, Delft, The Netherlands}
\author{Makars \v{S}i\v{s}kins}
\thanks{These authors contributed equally.}
\affiliation{Kavli Institute of Nanoscience, Delft University of Technology, Lorentzweg 1,\\ 2628 CJ, Delft, The Netherlands}
\author{Martin Lee}
\affiliation{Kavli Institute of Nanoscience, Delft University of Technology, Lorentzweg 1,\\ 2628 CJ, Delft, The Netherlands}

\author{Dorye L. Esteras}
\affiliation{Instituto de Ciencia Molecular (ICMol), Universitat de Val\`{e}ncia, c/Catedr\'{a}tico Jos\'{e} Beltr\'{a}n 2,\\ 46980 Paterna, Spain}
\author{Alberto M. Ruiz}
\affiliation{Instituto de Ciencia Molecular (ICMol), Universitat de Val\`{e}ncia, c/Catedr\'{a}tico Jos\'{e} Beltr\'{a}n 2,\\ 46980 Paterna, Spain}
\author{Samuel Ma\~{n}as-Valero}
\affiliation{Kavli Institute of Nanoscience, Delft University of Technology, Lorentzweg 1,\\ 2628 CJ, Delft, The Netherlands}
\affiliation{Instituto de Ciencia Molecular (ICMol), Universitat de Val\`{e}ncia, c/Catedr\'{a}tico Jos\'{e} Beltr\'{a}n 2,\\ 46980 Paterna, Spain}

\author{Carla Boix-Constant}
\affiliation{Instituto de Ciencia Molecular (ICMol), Universitat de Val\`{e}ncia, c/Catedr\'{a}tico Jos\'{e} Beltr\'{a}n 2,\\ 46980 Paterna, Spain}
\author{Jose J. Baldov\'{i}}
\affiliation{Instituto de Ciencia Molecular (ICMol), Universitat de Val\`{e}ncia, c/Catedr\'{a}tico Jos\'{e} Beltr\'{a}n 2,\\ 46980 Paterna, Spain}
\author{Eugenio Coronado}
\affiliation{Instituto de Ciencia Molecular (ICMol), Universitat de Val\`{e}ncia, c/Catedr\'{a}tico Jos\'{e} Beltr\'{a}n 2,\\ 46980 Paterna, Spain}
\author{Yaroslav M. Blanter}
\affiliation{Kavli Institute of Nanoscience, Delft University of Technology, Lorentzweg 1,\\ 2628 CJ, Delft, The Netherlands}
\author{Peter G. Steeneken}
\affiliation{Kavli Institute of Nanoscience, Delft University of Technology, Lorentzweg 1,\\ 2628 CJ, Delft, The Netherlands}
\affiliation{Department of Precision and Microsystems Engineering, Delft University of Technology,
	Mekelweg 2,\\ 2628 CD, Delft, The Netherlands}
\author{Herre S. J. van der Zant}
\affiliation{Kavli Institute of Nanoscience, Delft University of Technology, Lorentzweg 1,\\ 2628 CJ, Delft, The Netherlands}

\email{e-mail: makars@nus.edu.sg; m.j.a.houmes@tudelft.nl; h.s.j.vanderzant@tudelft.nl; p.g.steeneken@tudelft.nl}

\begin{abstract}
    The temperature dependent order parameter provides important information on the nature of magnetism. Using traditional methods to study this parameter in two-dimensional (2D) magnets remains difficult, however, particularly for insulating antiferromagnetic (AF) compounds. Here, we show that its temperature dependence in AF MPS$_{3}$ (M(II) = Fe, Co, Ni) can be probed via the anisotropy in the resonance frequency of rectangular membranes, mediated by a combination of anisotropic magnetostriction and spontaneous staggered magnetization. Density functional calculations followed by a derived orbital-resolved magnetic exchange analysis confirm and unravel the microscopic origin of this magnetization inducing anistropic strain. We further show that the temperature and thickness dependent order parameter allows to deduce the material’s critical exponents characterising magnetic order. Nanomechanical sensing of magnetic order thus provides a future platform to investigate 2D magnetism down to the single-layer limit.
\end{abstract}

\maketitle
Layered two-dimensional (2D) magnetic materials offer an emerging platform for fundamental studies of magnetism in the 2D limit. Their stackability into van der Waals heterostructures opens pathways to non-trivial magnetic phases and technological applications, including sensors, memories and spintronic logic devices~\cite{Mak2019}.
In addition to ferromagnetism, first observed in CrI$_3$~\cite{Huang2017} and Cr$_2$Ge$_2$Te$_6$~\cite{DiscoveryGong2017}, antiferromagnetism in 2D materials has also been studied in FePS$_{3}$~\cite{Lee2016} and CrSBr~\cite{Telford2020}.
Antiferromagnetic (AF) materials are of particular technological interest due to their high spin-wave propagation speed and lack of macroscopic stray fields, making them strong candidates for spintronic and magnonic applications~\cite{Nemec2018,Rahman2021,Mertens2022,Boix-Constant2022,Esteras2022}.

For insulating, thin AF materials, such as MPS$_{3}$ (M(II) = Fe, Co, Ni), few methods are available to study their intrinsic magnetism. Conventional techniques, such as neutron scattering, magnetization measurement by a superconducting quantum interference device (SQUID) or vibrating sample magnetometry are challenging, due to the small volumes of exfoliated 2D materials. Other methods, suited to 2D materials, require electrical conductance, the presence of specific optical modes or ferromagnetic order; they are therefore difficult to apply~\cite{Mak2019}. In contrast, strain applied to 2D magnetic materials was shown to be extremely powerful, inducing magnetization reversal~\cite{FengStrainReversal}, reorientating the easy-axis~\cite{NeelVectorNi2021}, or reversing the exchange interaction~\cite{cenker_reversible_2022}.
In addition, the direct coupling between strain, resonance frequency and magnetization in membranes of 2D magnets, makes nanomechanical resonance a sensitive method for studying their phase transitions~\cite{Siskins2020, siskins_nanomechanical_2022, FaiMakJiang2020}.
Here, we show, guided by density functional theory (DFT), that the magnetic order parameter of MPS$_3$ AF membranes can be quantified through the anisotropy in their magneto-elastic response; from its temperature dependence the critical exponents are determined, and their thickness dependence is investigated.

\begin{figure*}[ht]
\centering
	\includegraphics[width = \linewidth]{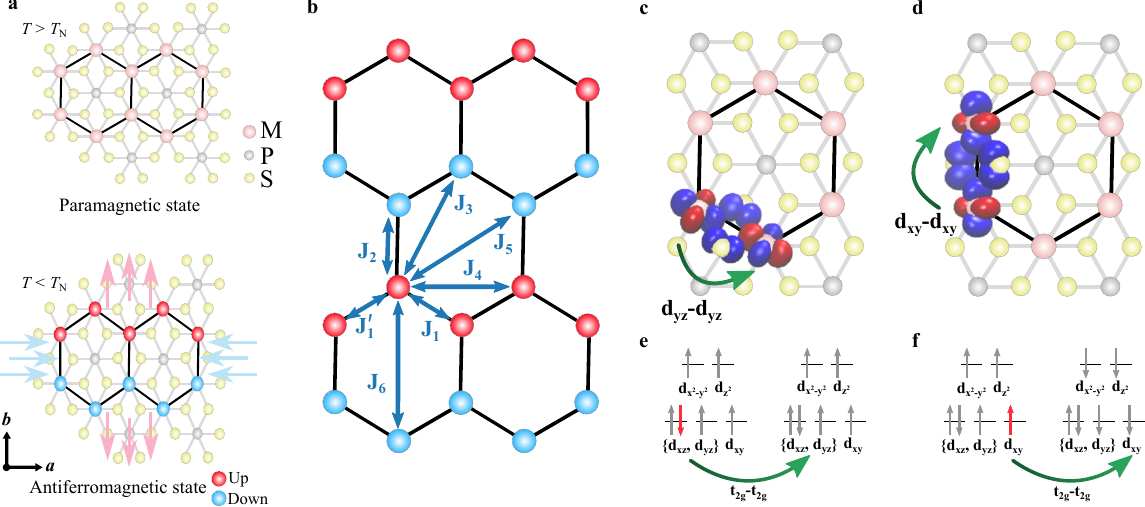}
	\caption{\textbf{Magnetostriction in MPS$_3$ membranes.} \textbf{a, top panel,} Crystalline structure of MPS$_3$ in the paramagnetic phase ($T>T_{\rm N}$). Black hexagons indicate the organisation of magnetic atoms in the lattice. \textbf{a, bottom panel,} Crystalline structure of MPS$_3$ at the AF phase ($T<T_{\rm N}$) as it elongates in the $b$ and contracts in the $a$ direction. Light blue and red arrows indicate the axial lattice distortion. \textbf{b,} Illustration of the exchange interaction parameters included into the Heisenberg spin Hamiltonian. \textbf{c-d,} Calculated maximally localized Wannier orbitals. Green arrows illustrate the most relevant FM superexchange channels for J$_{1}$ (J$^{'}_{1}$) (\textbf{c}) and J$_{2}$ (\textbf{d}), corresponding with the d$_{yz}$-d$_{yz}$ (d$_{xz}$-d$_{xz}$) and d$_{xy}$-d$_{xy}$ orbitals, respectively. \textbf{e-f,} Electron configuration of the Fe$^{2+}$ magnetic ions connected by J$_{1}$ (\textbf{e}) and J$_{2}$ (\textbf{f}), showing parallel and antiparallel spin orientations, respectively.}
	\label{fig:Fig1}
\end{figure*}

\section*{Results and Discussion}
\subsection*{First principles analysis of spontaneous magnetostriction in MPS$_{3}$}
Transition-metal phosphorus trisulphides, with general formula MPS$_3$, are layered materials stacked in a monoclinic lattice with symmetry group C2/m~\cite{Chittari2016}, as shown in the top view of a single-layer in the paramagnetic phase, Fig.~\ref{fig:Fig1}a, top panel. The spins of FePS$_3$ point out-of-plane, whereas both CoPS$_3$ and NiPS$_3$ are in-plane systems with their spins preferentially aligned along the $a$ axis.  The intralayer AF order forms a zigzag configuration, as shown in bottom panel of Fig.~\ref{fig:Fig1}a, leading to two opposite aligned magnetic sub-latices. The difference of the magnetisation between these sub-latices is the N\'{e}el vector.
In bulk CoPS$_3$ and NiPS$_3$, these layers with this staggered magnetism are stacked in a ferromagnetic (FM) fashion with N\'{e}el transition temperatures, $T_{\text{N}}$, around 119 and 155~K, respectively~\cite{Wildes2017,Joy1992}. The interlayer magnetic interactions in FePS$_3$ are AF with a transition around 118~K~\cite{Takano2004}. 

To analyse the effect of magnetic ordering on the lattice, we performed first principles structural optimizations of FePS$_{3}$, CoPS$_{3}$ and NiPS$_{3}$ based on density functional theory (DFT).
For the ground state zigzag magnetic configuration, the calculations predict a compression of the $a$ lattice parameter with respect to the crystallographic, non-magnetic structure of 2.545\% and 1.328\% for the Co and Fe derivatives respectively (see Table~\ref{table1}). 
In addition, the $b$ axis expands by 0.402\% (Co) and 0.359\% (Fe).
In contrast, in NiPS$_{3}$ the lattice parameters remain almost unchanged.
The crystal and magnetic structures are strongly connected in these compounds, which is further corroborated by simulations of different spin configurations (see Supplementary Note~1).

The microscopic mechanism governing the spontaneous magnetostriction in these materials is studied using orbital-resolved magnetic exchange analyses based on maximally localized Wannier functions, (see Supplementary Note~1). The analysis shows that the spontaneous magnetostriction calculated in FePS$_{3}$ and CoPS$_{3}$ arises from isotropic magnetic exchange interactions between t$_{2g}$-t$_{2g}$ orbitals.
Specifically, for FePS$_{3}$ the main magnetic exchange channels, substantially affected by the compression of the $a$ and expansion of the $b$ lattice parameters, are the ones involving t$_{2g}$-t$_{2g}$ interactions of FM nature. The changes in the lattice parameters result in an increase in J$_{1}$ and J$'_{1}$ due to a decrease in distance between the d$_{yz}$-d$_{yz}$ and d$_{xz}$-d$_{xz}$ orbitals, respectively (Fig.~\ref{fig:Fig1}c).
Simultaneously, these changes cause a decrease of J$_{2}$ due to a larger separation of the d$_{xy}$-d$_{xy}$ orbitals (Fig.~\ref{fig:Fig1}d).
This is compatible with the electron configuration of Fe$^{2+}$ (d$^{6}$), which has these orbitals partially filled and allows FM hopping between them (Fig.~\ref{fig:Fig1}e,f).

This hopping effect also occurs for Co$^{2+}$ (d$^{7}$) although the additional electron present for Co blocks the d$_{xy}$-d$_{xy}$ pathway (Supplementary Note~1, Fig.~S2).
This results in a stronger effect along J$_{1}$ and J$'_{1}$ for the optimized structure, maximizing FM interactions in the zigzag chain, which involve the d$_{yz}$-d$_{yz}$ and d$_{xz}$-d$_{xz}$ orbitals, respectively.
For the Ni$^{2+}$ derivative (d$^{8}$), the t$_{2g}$ energy levels are fully occupied (Supplementary Note~1, Fig.~S3), which results in a blocking of the t$_{2g}$-t$_{2g}$ magnetic super-exchange channels. This leads to an almost negligible modification in the lattice parameters of the optimized structure with respect to the crystallographic non-magnetic one. 

{\renewcommand{\arraystretch}{1.5}
\begin{table*}[]
\caption{CoPS$_3$, FePS$_3$ and NiPS$_3$ lattice parameters of the crystallographic non-magnetic (NM) and fully optimized zigzag antiferromagnetic (AF-zigzag) configurations, as calculated by DFT (see Supplementary Note~1).}
\begin{tabular}{ccccccc}
& \multicolumn{2}{c}{\textbf{CoPS$_{3}$}}        & \multicolumn{2}{c}{\textbf{FePS$_{3}$}}        & \multicolumn{2}{c}{\textbf{NiPS$_{3}$}}        \\ \hline
\textbf{Lattice parameter (\textbf{\AA}) } & \textit{\textbf{a}} & \textit{\textbf{b}} & \textit{\textbf{a}} & \textit{\textbf{b}} & \textit{\textbf{a}} & \textit{\textbf{b}} \\
\textbf{NM}                & 5.895               & 10.19               & 5.947               & 10.301              & 5.812               & 10.07               \\
\textbf{AF-zigzag}      & 5.745               & 10.231              & 5.868               & 10.338              & 5.817               & 10.061              \\ \hline
{\textbf{Change ($\%$)}} & -2.545            & +0.402            & -1.328           & +0.359            & +0.086            & -0.089           
\end{tabular}
\label{table1}
\end{table*}}

\subsection*{Resonance frequency changes due to spontaneous magnetostrictive strain}\label{sub sec: Spontaneous magnetostrictive strain}
\begin{figure*}[ht]
\centering
	\includegraphics[width = \linewidth]{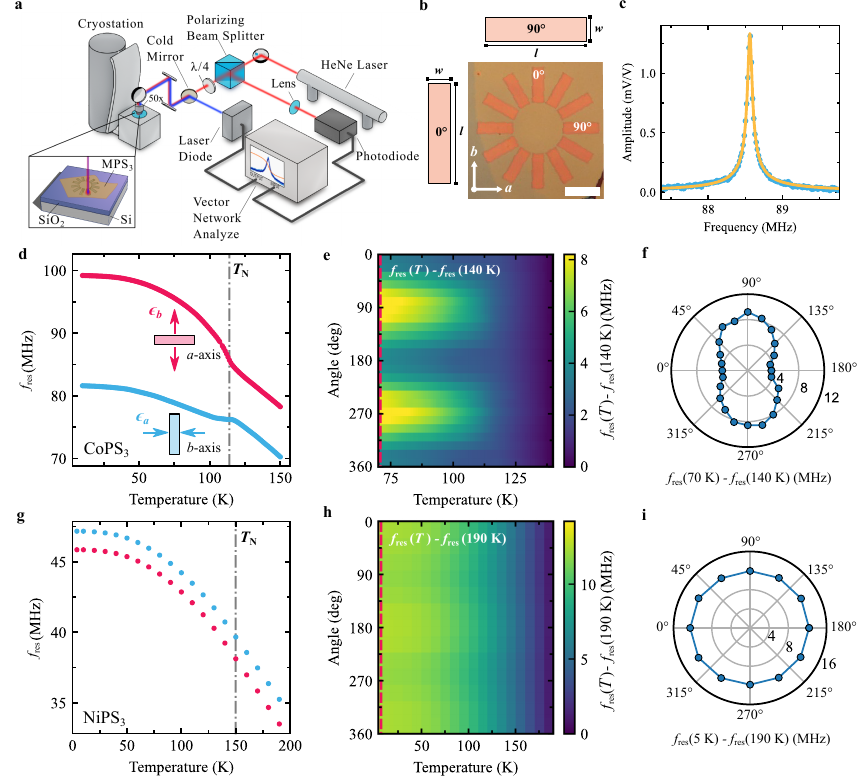}
 	\caption{\textbf{Angle-resolved mechanical characterization via laser interferometry.} \textbf{a,} Schematic illustration of the laser interferometry setup and sample with rectangular cavity array. \textbf{b,} Optical image of the rectangular membranes array for a CoPS$_{3}$ sample. The $a$ and $b$ axis are determined from the resonance frequency behaviour. Scale bar: $12\, \mu$m.  Schematic of $0^\circ$ and $90^\circ$ membranes from the array where $w$ is the width of the membrane and $l$ its length. \textbf{c,} Measured amplitude of the fundamental resonance peak in a CoPS$_{3}$ drum at $T = 10$ K and Lorentzian fit used to extract the fundamental resonance frequency, $f_{\text{res}}$, and quality factor, $Q$. \textbf{d,} Temperature dependence of $f_{\text{res}}$ of a CoPS$_3$ rectangular membrane orientated along the $a$ and $b$ axes. The dashed line indicates the transition temperature $T_{\rm N}$ extracted from the data. \textbf{e,} Resonance frequency difference, $f_{\text{res}}(T)-f_{\text{res}}(140\,\mathrm{K})$, as a function of angle and temperature. The dashed line indicates the transition as in \textbf{d} \textbf{f,} Polar plot of $f_{\text{res}}(T)-f_{\text{res}}(140\,\mathrm{K})$ taken along the red dashed line in (e). Panels \textbf{g-i,} follow the same structure as (c-e) for NiPS$_3$ resonators with negligible anisotropy, measured between 5 K and 190 K.}
	\label{fig:Fig2}
\end{figure*}

The predicted anisotropic change of lattice parameters when going from the paramagnetic to the AF phase, causes compressive stress, $\sigma_{\text{a}}$, and tensile stress, $\sigma_{\text{b}}$, along the \textit{a} axis and \textit{b} axis respectively, as illustrated in Fig.~\ref{fig:Fig1}a, bottom pannel. To quantify this anisotropy appearing at the phase transition, we use rectangular membranes, shown in Fig.~\ref{fig:Fig2}b, to nanomechanically probe stress variations, along a specific crystallographic axis~\cite{SiskinsAs2S32019} (see Supplementary Note~2).
In the following analysis, we neglect the stress contribution from the thermal expansion of the substrate, as this is small compared to that of the MPS$_{3}$ compounds \cite{Siskins2020}.

The resonance frequency of the fundamental mode of a rectangular membrane, $f_\text{res}$, is approximately given by~\cite{Leissa2011}:
\begin{equation}
    \label{eq:rectangular_membrane_freq}
    f_\text{res} \approx \frac{1}{2}\sqrt{\frac{1}{\rho }\left[\frac{1}{w^2}\sigma_{\text{w}}+\frac{1}{l^2}\sigma_{\text{l}}\right]}\,,
\end{equation}
where $\rho$ is the mass density, $w$ and $l$ are respectively the width and length of the membrane, as indicated in Fig.~\ref{fig:Fig2}b, and $\sigma_{\text{w,l}}$ are the stresses parallel to these directions. For high-aspect-ratio membranes ($w\ll l$), the mechanical resonance frequency is mostly determined by the stress along the shortest direction, $\sigma_{\text{w}}$. 

We study the resonance frequency of thin MPS$_3$ flakes suspended over star-shaped cavities with $30^{\circ}$ angular resolution, as shown in an example device in Fig.~ \ref{fig:Fig2}b. When the longest side of the cavity is aligned along a crystallographic axis (\textit{a} or \textit{b}) and $w\ll l$, its fundamental resonance frequency ($f_\text{a}$ or $f_\text{b}$) is determined by the stress along the perpendicular axis ($\sigma_{\text{b}}$ or $\sigma_{\text{a}}$):
\begin{equation}
    \label{eq:freq_0deg_90deg}
     f_{\text{a}}\approx\frac{1}{2}\sqrt{\frac{1}{\rho  w^2}\sigma_\text{b}} ~\text{and}~  f_{\text{b}}\approx\frac{1}{2}\sqrt{\frac{1}{\rho w^2}\sigma_\text{a}}\, .
\end{equation} 
On cavities oriented at an intermediate angle, $\theta$, (defined with respect to the $b$ axis), the resonance frequency is:
\begin{eqnarray}\label{eq:freq_theta deg}
    f_\theta(T) &\approx &\frac{1}{2}\sqrt{\frac{1}{\rho w^2}\Big[\sigma_{\text{a},\theta} +\sigma_{\text{b},\theta} \Big]},\\
    \sigma_{\text{a},\theta} &=& \frac{E}{(1-\nu^2)}(\cos^2\theta+\nu\sin^2\theta)(\bar{\epsilon}-\epsilon_{\text{ms},\text{a}}), \nonumber \\
    \sigma_{\text{b},\theta} &=& \frac{E}{(1-\nu^2)} (\sin^2\theta+\nu\cos^2\theta)(\bar{\epsilon}-\epsilon_{\text{ms},\text{b}}) ,\nonumber 
\end{eqnarray}
where we have used the constitutive equations for a magnetostrictive membrane with plane stress~\cite{Landau1984}, while only keeping the anisotropy in the magnetostriction coefficient, Supplementary Note~2. Here, $E$ is the Young's modulus and $\nu$ is Poisson's ratio of the material. Moreover, we have
$\bar{\epsilon} = \epsilon_{\text{fab}} - \epsilon_{\text{th}}$, with  $\epsilon_{\text{fab}}$ the residual fabrication strain and $\epsilon_{\text{th}}$ the phononic thermal expansion induced strain variation. The magnetostrictive strain along the $a$ and $b$-axes is given by $\epsilon_{\text{ms},\text{a,b}}=\lambda_{\text{a,b}} L^2$, respectively (see Supplementary Note~3 for a detailed derivation of Eq.~\eqref{eq:freq_theta deg}), where $\lambda_{\text{a,b}}$ are magnetostriction coefficients and $L^{2}$ is the AF order parameter squared.

The temperature dependence of the resonance frequency comprises two contributions: one due to the phononic thermal expansion coefficient $\alpha$, given by
$\epsilon_{\text{th}}(T) = \int_{T_{0}}^{T} \alpha(\tilde{T}) \rm{d}\tilde{T}$, where $T_0$ is a reference temperature and $\tilde{T}$ the integration variable, and the magnetostrictive contribution $\epsilon_{\text{ms},\text{a,b}}(T)=\lambda_{\text{a,b}} L^2(T)$. The former contribution is a slowly varying function of $T$, while the latter term contains the staggered magnetization, which increases abruptly near the phase transition; it thus can be used to determine $L(T)$, as we will show below. We assume $\lambda_{\text{a,b}}$ to be $T$ independent, as its temperature dependence will be negligible when compared to that of $L(T)$.

\subsection*{Nanomechanical determination of the order parameter}
To quantify the anisotropy in the magnetic membranes, a laser interferometry technique is used to measure their resonance frequency as a function of temperature~\cite{Bunch490}.
A MPS$_3$ flake, suspended over holes in a patterned Si/SiO$_2$ chip, Fig.~\ref{fig:Fig2}b, is placed inside a cryostat with optical access as shown in Fig.~\ref{fig:Fig2}a.
Both actuation and detection are done optically, by means of a power-modulated blue laser which opto-thermally excites the fundamental resonance, and a constant red laser which measures the change in the reflected signal resulting from the membrane's motion~\cite{Siskins2020}. A typical resonance is shown in Fig.~\ref{fig:Fig2}c, along with the damped harmonic oscillator model fit defining the resonance frequency.
Figure \ref{fig:Fig2}d shows that in CoPS$_{3}$  $f_{\text{a}}$ and $f_{\text{b}}$ exhibit a similar temperature dependence for $T>T_{\rm N}$, while opposite behaviour below the phase transition is visible, namely an increase of $f_{\text{a}}$ and a relative decrease of $f_{\text{b}}$. This sudden change in $f(T)$ for the perpendicular cavities, occurring near $T_{\text{N}}$, constitutes, in accordance with the DFT calculations, the central result of this work as it shows that the magnetic ordering in MPS$_3$ leads to anisotropic strain and thus spontaneous magnetostriction. We further note that strictly speaking, $T_{\rm N}$ should be replaced by $T_{\rm N}^*$ which includes the effects of strain (see Supplementary Note~2). For simplicity, we here use the notation $T_{\rm N}$ for the measured transition temperatures.  

The anisotropic behavior of CoPS$_3$ in the AF state is even more evident in Fig.~\ref{fig:Fig2}e, where $f_{\text{res}}(T) - f_{\text{res}}(140 \text{ K})$ for the different cavities of the star-shaped sample are plotted as a function of $\theta$ and temperature. The polar plot in Fig.~\ref{fig:Fig2}f shows the data along the red dashed line at $T = 70$ K in Fig.~\ref{fig:Fig2}e and results in a characteristic dumbbell-shape. Similar anisotropic behaviour is observed in FePS$_3$ as shown in Supplementary Note~4. On the contrary, for NiPS$_{3}$ negligible anisotropy is observed in the angle-resolved magnetostriction data in Fig.~\ref{fig:Fig2}g-i. 

\begin{figure*}[ht]\centering
	\includegraphics[width=\linewidth]{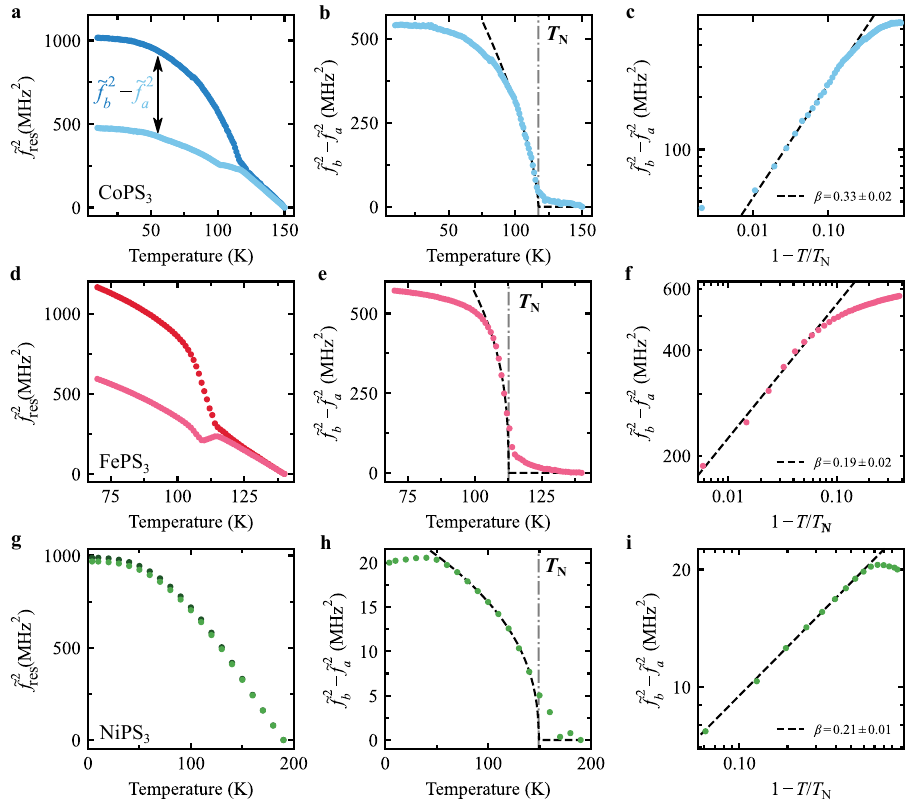}
   	\caption{\textbf{Anisotropy and critical behaviour in resonance frequency of MPS$_3$ (M(II) = Co, Fe, Ni) membranes.} \textbf{a,} Pretension corrected resonance frequency ($\tilde{f}^2_{\text{a}}(T) = f_{\text{a}}^2(T) - f_{\text{a}}^2(150\text{K})$ and $\tilde{f}^2_{\text{b}}(T) = f_{\text{b}}^2(T) - f_{\text{b}}^2(150\text{K})$) of rectangular membranes of CoPS$_3$ oriented along the $b$ axis (blue) and $a$ axis (red) \textbf{b,} Difference of the corrected frequency squared $\tilde{f}_{\text{b}}^2 - \tilde{f}_{\text{a}}^2$ proportional to the order parameter $L^2$ from Eq. \ref{eq:freq_diff_order}. The dashed-dotted line indicates the measured transition temperature $T_{\rm N}$. The dashed black line is a powerlaw fit through the data close to $T_{\rm N}$ (see Supplementary Note~6). \textbf{c,} Difference of the corrected frequency squared $\tilde{f}_{\text{b}}^2 - \tilde{f}_{\text{a}}^2$ as a function of the reduced temperature $1-T/T_{\rm N}$. The dashed black line is the fit from b where the slope defines the critical exponent $2\beta$. \textbf{d-f,} and \textbf{g-i,} follow the same structure as (a-d) for FePS$_3$ and NiPS$_3$ resonators, respectively.}
	\label{fig:Fig3}
\end{figure*}

To obtain $L(T)$ from the data, we first subtract the pretension contribution from the resonance frequency $f_{\theta}(T_0)$ by calculating $\tilde{f}_{\theta}^2(T) =f_{\theta}^2(T) -f_{\theta}^2(T_0)$, for each angle, where $T_0 = 150$~K is the highest temperature in our measurements. The resulting values of $\tilde{f}_{\theta}^2(T)$ along the crystalline axes $a$ and $b$ are shown in Fig.~\ref{fig:Fig3}a,d,g for the three MPS$_3$ compounds. With Eq. \eqref{eq:freq_theta deg}, we then calculate the difference  $\tilde{f}_{\text{b}}^2(T) - \tilde{f}_{\text{a}}^2(T)$ which yields
\begin{align}\label{eq:freq_diff_order}
    \tilde{f}^2_{\text{b}}-\tilde{f}^2_{\text{a}}=\frac{E}{4 \rho w^2(1+\nu)}\left[\lambda_{\text{a}}-\lambda_{\text{b}}\right]L^2.
\end{align}
We can now use Eq.~\eqref{eq:freq_diff_order} to access the critical behaviour of $L$ below $T_{\rm N}$ by plotting $\tilde{f}_{{\text{b}}}^2 -\tilde{f}_{\text{a}}^2$ as a function of temperature. As shown in  Fig.~\ref{fig:Fig3}b,e,h, the trend presents the typical critical behaviour with a non-zero order parameter appearing in the ordered state for $T<T_{\rm N}$. Figures~\ref{fig:Fig3}c,f,i show the same critical curve as Fig.~\ref{fig:Fig3}b,e,h respectively, plotted on a logarithmic scale against the reduced temperature $(1-T/T_\text{N})$. 
Note that the difference $\tilde{f}_{{\text{b}}}^2 -\tilde{f}_{\text{a}}^2$ for NiPS$_3$, is substantially smaller than that of the Fe/CoPS$_3$ membranes indicative of a weaker anisotropic magnetostrictive behaviour.

The angle dependence of the resonance frequencies allows us to estimate the ratio $r_{\text{ab}}=\lambda_{\text{a}}/\lambda_{\text{b}}$ between the magnetostriction parameters, $\lambda_{\text{a,b}}$, (see Supplementary Note~3). This ratio we directly compare to DFT calculations:
Experimentally, we find for FePS$_3$,  $r_{\text{ab}}=-2.3\pm0.3$ while from the DFT calculations we estimate $r_{\text{ab}}=-3.70$.
For CoPS$_{3}$ (taking~\cite{Gui2021} $\nu_{\text{CoPS}_3}=0.293$), the experimental value is $-1.42\pm0.07$ and the DFT one $-6.33$. We conclude that although both the sign and order of magnitude of the magnetostrictive anisotropy in these compounds are well reproduced in the current work, more detailed studies will be needed to obtain full quantitative correspondence with theory.

\begin{figure}\centering
	\includegraphics[width=0.9\linewidth]{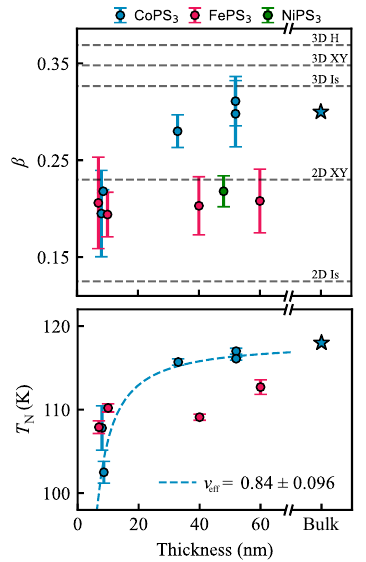}
    \caption{\textbf{Thickness dependence of  critical behaviour.} 
    Average critical exponent, $\beta$, and critical temperature, $T_{\rm N}$, of MPS$_3$ resonators plotted as a function of thickness. The blue stars indicate CoPS$_3$ bulk values from~\cite{QiyeLiu2021}. Critical parameters have been determined  from power law fits to $\tilde{f}_{\text{b}}^2 - \tilde{f}_{\theta}^2$, as shown in Fig.~\ref{fig:Fig3}b,e,h, and then taking the average value over the fit parameter for all angles $\theta \neq 0$. Error bars are calculated from standard deviation of fit results for all $\theta$. The horizontal gray dashed lines in the upper plot indicate the expected values of $\beta$ for the 3D or 2D versions of the Heisenberg (H), XY or Ising (Is) models. The blue dashed line in the lowe panel indicates a fit to Eq.~\eqref{eq: thickness powerlaw} through the CoPS$_3$ data with $\nu_\text{eff} =0.84\pm0.13$.}
	\label{fig:Fig4}
\end{figure}

\subsection*{Thickness dependence of critical behaviour}\label{sub sec: Materials with weak magnetostriction}
As follows from Landau's theory of phase transitions (see Supplementary Note~2), $L(T)$ near $T_{\rm N}$ is given by
\begin{equation} \label{eq: L^2 powerlaw}
    L^2(T) = \begin{cases}
    0 & \text{if } T>T_{\rm{N}} \\
    \frac{A}{2B}(T_{\rm N}-T)^{2\beta} & \text{if } T<T_{\rm{N}},
    \end{cases}
\end{equation}
where $A$ and $B$ are constants and $\beta$ is a critical exponent representative of the magnetic order. We fit Eq.~\eqref{eq: L^2 powerlaw} to the data in Fig.~\ref{fig:Fig3}b,e,h in the region close to $T_\text{N}$ (indicated by the black dashed line in Fig.~\ref{fig:Fig3}b,e,h) to extract the critical exponent $\beta$ and $T_\text{N}$ for the three materials (see Supplementary Note~6 for more details on the fitting procedure). In the logarithmic plot of the critical curve the fitting of a straight line shows good agreement to the data points, consistent with the result of Eq.~\eqref{eq: L^2 powerlaw}.
The values for $\beta$ and $T_\text{N}$ are plotted in Fig.~\ref{fig:Fig4} as a function of thickness, $t$, and listed in Supplementary Note~6, Table 49.

For the weakly anisotropic NiPS$_3$, $\beta = 0.218\pm0.016$, comparable to the value ($\beta=0.22\pm0.02$) found in Ref.~\cite{Afanasiev2020}, and consistent with the expected 2D XY magnetic dimensionality ($\beta_\text{2DXY} = 0.233$) of NiPS$_3$ \cite{Kim2019}. 
For FePS$_3$ we find $\beta = 0.208 \pm 0.033$,  comparable with literature values~\cite{yao2004mossbauer}. For both $\beta$ and $T_{\text{N}}$ no appreciable thickness dependence is observed, similar to what has previously been reported in Ref.~\cite{zhang_observation_2021}, where changes in the critical behaviour mostly become visible in the monolayer limit.

For thicker CoPS$_3$ samples ($t = 40-60$~nm) we find $\beta =0.289 \pm 0.034$ close to what is reported in literature for the bulk ($\beta_\text{bulk} = 0.3\pm 0.01$~\cite{Wildes2017}) and consistent with the 3D Ising model. For samples with $t< 10$~nm the measured $\beta$, on the other hand, is $0.195 \pm 0.045$, closer to $\beta_\text{2DXY}$ as shown in the top panel of Fig.~\ref{fig:Fig4}. This constitutes a noticeable change in $\beta$ while going from bulk to thinner samples. Similarly, we observe for CoPS$_3$ a decrease in $T_{\rm N}$ from the bulk value of $118$~K down to $\sim 100$~K, similar to what was previously reported in Ref.~\cite{QiyeLiu2021}.
We fit a power law to the dependence of $T_{\rm N}$ on thickness,
\begin{equation}\label{eq: thickness powerlaw}
    T_\text{N}(t)/T_\text{N}^\text{3D} \propto 1-(C/t)^{1/\nu_\text{eff}},
\end{equation}
where $C$ is a non-universal constant related to the interlayer coupling, and $\nu_\text{eff}$ is an effective critical exponent related to the correlation length~\cite{Zhang2001}. Fitting the CoPS$_3$ data points with $T_\text{N}^\text{3D} = 118$~K~\cite{QiyeLiu2021} yields $C = 1.43 \pm 0.457$~nm and $\nu_\text{eff} = 0.84\pm 0.096$. This value of $\nu_\text{eff}$ is intermediate between the expected values of $\nu_\text{eff} = 0.630$ for the 3D Ising and $\nu_\text{eff} = 1$ for the 2D Ising models, and indicative of a transition regime~\cite{Gibertini2019}.

\section*{Conclusions}
In conclusion, we provide a comprehensive analysis of the anisotropic magnetostriction effect in MPS$_3$ compounds and its implications to the dynamics of membrane made from them.
DFT calculations provide a microscopic explanation for the anisotropic lattice deformation in CoPS$_3$, FePS$_3$ and NiPS$_3$ which are consistent with our measurements.
We further demonstrate the relation between magnetic ordering and anisotropy in the mechanical resonance frequency of suspended MPS$_3$ resonators, providing a direct measure of the AF  order parameter in absence of an external magnetic field.
We observe a thickness dependence in the critical behaviour of CoPS$_3$ resonators~\cite{Wildes2017, QiyeLiu2021}, which is absent in the case of FePS$_3$. 
The presented technique is of particular interest for the study of 2D magnetism given the scarcity of methods available to investigate critical phenomena  of van der Waals materials in the atomically thin limit.

\section*{Methods}\label{methods}
\noindent\textbf{Sample fabrication.}
Substrates consist of thermal SiO$_2$ of 285 nm thickness, grown on highly doped (Si$^{++}$) silicon. The rectangular cavities are defined via e-beam lithography using AR-P 6200 resist. After development, the exposed SiO$_2$ areas are fully etched via reactive ion etching. The AR-P 6200 resist is stripped in PRS-3000 and the sample is cleaned in an O$_2$ plasma before stamping. The exfoliation and transfer of multi-layer MPS$_3$ flakes is done using a polydimethylsiloxaan (PDMS) transfer method. First, MPS$_3$ crystals are exfoliated onto the PDMS through scotch tape. Selected flakes are then transferred on the star-shaped cavities in the SiO$_2$/Si substrate.
\newline\newline
\textbf{Laser interferometry.} Samples are mounted on a heater stage which is cooled down to 4~K using a Montana Instruments Cryostation s50 cryostat with optical access. A blue diode laser ($\lambda = 405$~nm) is used to excite the membrane optothermally via AC power-modulation from a vector network analyzer (VNA)~\cite{Steeneken_2021}. Displacements are detected by focusing a red He-Ne laser beam ($\lambda = 632$~nm) on the cavity formed by the membrane and Si substrate. The reflected light, which is modulated by the position-dependent membrane motion, is recorded by a photodiode and processed by a phase-sensitive VNA. Laser spot size is $\sim 1 \,\mu$m. 
\newline\newline
\textbf{DFT calculations.}
First principles spin-polarized DFT calculations in the plane wave formalism are preformed as implemented in the Quantum ESPRESSO package~\cite{QuanEssprGiannozzi2009}. The exchange-correlation energy is calculated using the generalized gradient approximation using the Perdew–Burke–Ernzerhof functional~\cite{Perdew1996} and standard Ultra-soft (USPP) solid-state pseudopotentials. The electronic wave functions are expanded with well-converged kinetic energy cut-offs for the wave functions (charge density) of $75$ ($800$), $85$ ($800$) and $85$ ($800$) Ry for Fe, Co and Ni, respectively. The crystal structures are fully optimized using the Broyden-Fletcher-Goldfarb-Shanno (BFGS) algorithm~\cite{Head1985} until the forces on each atom are smaller than $1\times10^{-3}$ Ry/au and the energy difference between two consecutive relaxation steps is less than $1\times10^{-4}$~Ry. In order to avoid unphysical interactions between images along the non-periodic direction, we add a vacuum of 18 \AA\, in the $z$ direction for the monolayer calculations. The Brillouin zone is sampled by a fine $\Gamma$-centered $5\times5\times1$ $k$-point Monkhorst–Pack~\cite{Monkhorst1976}. A tight-binding Hamiltonian derived from first-principles is constructed in the base of Maximally-localized Wannier functions, as implemented in the Wannier90 code~\cite{Mostofi2008}. For that, we select the d orbitals of the metal centre (Fe, Co, Ni) and the s and p orbitals of P and S to construct the connected subspace. Magnetic interactions are determined using the Green’s function method in the TB2J software~\cite{He2021}. The orbital resolved analysis is performed after rotating the coordinate system of the crystal to align the metal-sulfur bonds direction of the octahedra with the cartesian axes. 
\newline\newline
\textbf{Crystal growth}
Crystal growth of MPS$_3$ (M(II) = Ni, Fe, Co) is performed following a solid-state reaction inside a sealed evacuated quartz tube (pressure $\sim5\times10^{-5}$ mbar). I$_2$ was used as a transport agent to obtain large crystals. A three-zone furnace is used, where a tube with the material was placed in the leftmost zone. This side is then heated up to $700$ $^{\circ}$C in $3$ hours so that a temperature gradient of $700/ 650 / 675$~$^{\circ}$C is established. The other two zones are heated up in $24$ hours from room temperature to $650$~$^{\circ}$C and kept at that temperature for one day. The temperature is kept constant for $28$ days and cooled down naturally. With this process crystals with a length up to several centimeters are obtained. Detailed description of the crystal growth and characterization can be found in earlier work~\cite{Siskins2020}.  

\subsection*{Data availability}
All data supporting the findings of this article and its Supplementary Information
will be made available upon request to the authors.
\subsection*{Acknowledgments}
M.\v{S}., M.J.A.H., G.B., M.L., H.S.J.v.d.Z. and P.G.S. acknowledge funding from the European Union's Horizon 2020 research and innovation program under grant agreement number 881603. Y.M.B and H.S.J.v.d.Z. acknowledge support from Dutch National Science Foundation (NWO).
D.L.E., A.M.R., S.M.-V., C.B.-C., J.J.B., E.C. acknowledge funding from the European Union (ERC AdG Mol-2D 788222, ERC StG 2D-SMARTiES 101042680 and FET OPEN SINFONIA 964396), the Spanish MCIN (Project 2DHETEROS PID2020-117152RB-100 and Excellence Unit ”Maria de Maeztu”CEX2019-000919 -M), the Spanish MIU (FPU21/04195 to A.M.R.) and the Generalitat Valenciana (PROMETEO Program and APOST Grant CIAPOS/2021/215 to S.M.-V.) The computations were performed on the Tirant III cluster of the Servei d’Informàtica of the University of Valencia.
\subsection*{Author contributions}
D.L.E., A.M.R. performed the DFT and orbital resolved tight-binding calculations, supervised by J.J.B. M.\v{S}., M.J.A.H. and G.B. performed the laser interferometry measurements and fabricated and inspected the samples. M.L. and  M.J.A.H. fabricated the substrates. S.M.-V., C.B.-C. and E.C. synthesized and characterized the FePS$_{3}$, CoPS$_{3}$ and NiPS$_{3}$ crystals, supervised by E.c. M.\v{S}., M.J.A.H. and G.B. analysed the experimental data. M.\v{S}., M.J.A.H., Y.M.B., and P.G.S. modeled the experimental data. H.S.J.v.d.Z. and P.G.S. supervised the project. The paper was jointly written by all authors with a main contribution from  M.J.A.H. All authors discussed the results and commented on the paper.
\subsection*{Competing interests}
The authors declare no competing interests.

\newpage
\putbib

\end{document}



\title{SUPPLEMENTARY INFORMATION: Magnetic order in 2D antiferromagnets revealed by spontaneous anisotropic magnetostriction}

\affiliation{Kavli Institute of Nanoscience, Delft University of Technology, Lorentzweg 1,\\ 2628 CJ, Delft, The Netherlands}
\affiliation{Instituto de Ciencia Molecular (ICMol), Universitat de Val\`{e}ncia, c/Catedr\'{a}tico Jos\'{e} Beltr\'{a}n 2,\\ 46980 Paterna, Spain}
\affiliation{Department of Precision and Microsystems Engineering, Delft University of Technology,
	Mekelweg 2,\\ 2628 CD, Delft, The Netherlands}

\author{Maurits J. A. Houmes}
\thanks{These authors contributed equally.}
\affiliation{Kavli Institute of Nanoscience, Delft University of Technology, Lorentzweg 1,\\ 2628 CJ, Delft, The Netherlands}
\author{Gabriele Baglioni}
\thanks{These authors contributed equally.}
\affiliation{Kavli Institute of Nanoscience, Delft University of Technology, Lorentzweg 1,\\ 2628 CJ, Delft, The Netherlands}
\author{Makars \v{S}i\v{s}kins}
\thanks{These authors contributed equally.}
\affiliation{Kavli Institute of Nanoscience, Delft University of Technology, Lorentzweg 1,\\ 2628 CJ, Delft, The Netherlands}
\author{Martin Lee}
\affiliation{Kavli Institute of Nanoscience, Delft University of Technology, Lorentzweg 1,\\ 2628 CJ, Delft, The Netherlands}

\author{Dorye L. Esteras}
\affiliation{Instituto de Ciencia Molecular (ICMol), Universitat de Val\`{e}ncia, c/Catedr\'{a}tico Jos\'{e} Beltr\'{a}n 2,\\ 46980 Paterna, Spain}
\author{Alberto M. Ruiz}
\affiliation{Instituto de Ciencia Molecular (ICMol), Universitat de Val\`{e}ncia, c/Catedr\'{a}tico Jos\'{e} Beltr\'{a}n 2,\\ 46980 Paterna, Spain}
\author{Samuel Ma\~{n}as-Valero}
\affiliation{Kavli Institute of Nanoscience, Delft University of Technology, Lorentzweg 1,\\ 2628 CJ, Delft, The Netherlands}
\affiliation{Instituto de Ciencia Molecular (ICMol), Universitat de Val\`{e}ncia, c/Catedr\'{a}tico Jos\'{e} Beltr\'{a}n 2,\\ 46980 Paterna, Spain}

\author{Carla Boix-Constant}
\affiliation{Instituto de Ciencia Molecular (ICMol), Universitat de Val\`{e}ncia, c/Catedr\'{a}tico Jos\'{e} Beltr\'{a}n 2,\\ 46980 Paterna, Spain}
\author{Jose J. Baldov\'{i}}
\affiliation{Instituto de Ciencia Molecular (ICMol), Universitat de Val\`{e}ncia, c/Catedr\'{a}tico Jos\'{e} Beltr\'{a}n 2,\\ 46980 Paterna, Spain}
\author{Eugenio Coronado}
\affiliation{Instituto de Ciencia Molecular (ICMol), Universitat de Val\`{e}ncia, c/Catedr\'{a}tico Jos\'{e} Beltr\'{a}n 2,\\ 46980 Paterna, Spain}
\author{Yaroslav M. Blanter}
\affiliation{Kavli Institute of Nanoscience, Delft University of Technology, Lorentzweg 1,\\ 2628 CJ, Delft, The Netherlands}
\author{Peter G. Steeneken}
\affiliation{Kavli Institute of Nanoscience, Delft University of Technology, Lorentzweg 1,\\ 2628 CJ, Delft, The Netherlands}
\affiliation{Department of Precision and Microsystems Engineering, Delft University of Technology,
	Mekelweg 2,\\ 2628 CD, Delft, The Netherlands}
\author{Herre S. J. van der Zant}
\affiliation{Kavli Institute of Nanoscience, Delft University of Technology, Lorentzweg 1,\\ 2628 CJ, Delft, The Netherlands}

\email{e-mail: makars@nus.edu.sg; m.j.a.houmes@tudelft.nl; h.s.j.vanderzant@tudelft.nl; p.g.steeneken@tudelft.nl}

\maketitle
\onecolumngrid
\newpage
\section{Density Functional Theory calculations} 
This section summarizes the results of orbital-resolved magnetic exchange analyses based on maximally localized Wannier functions, for both the crystallographic and optimized structures. These results are included in Supplementary Tables~$7-42$. 
Supplementary Tables~$43-48$ display the hopping integrals with the S atoms (Supplementary Fig.~\ref{FigS4: Sulfur hopping}) that mediate the super-exchange interactions between magnetic centers for each channel. 
\begin{figure}[ht]
    \centering
    \includegraphics[width=.5\linewidth]{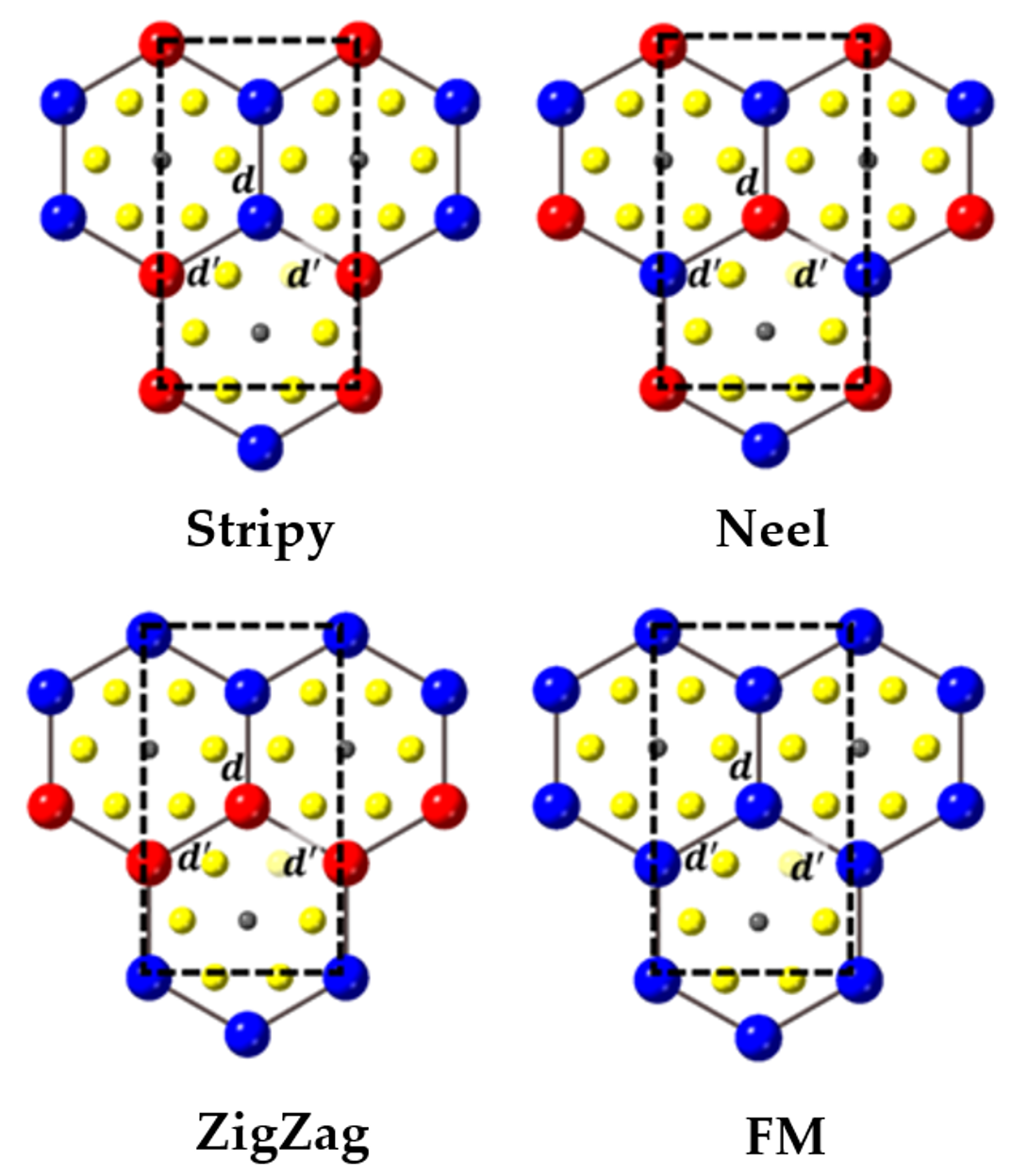}
    \caption{Top view of a single-layer of MPS$_{3}$ for different magnetic configurations, namely stripy, N\'{e}el, zigzag and ferromagnetic (FM). Blue and red balls represent transition metal atoms with spin up and down components, respectively.}
    \label{SImagconfig}
\end{figure}

\begin{figure}[ht]
    \centering
    \includegraphics[width=.7\linewidth]{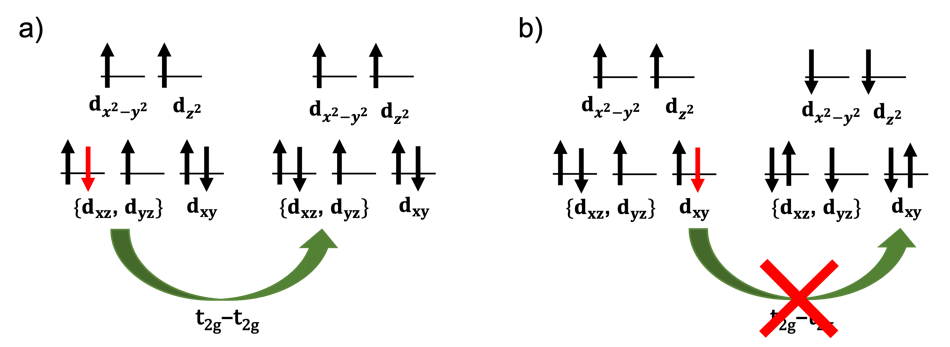}
    \caption{Electron configuration of the Co$^{2+}$ magnetic ions connected by J$_{1}$ (a) and J$_{2}$ (b), showing parallel and antiparallel spin orientations, respectively. Green arrows illustrate the most relevant ferromagnetic superexchange channels, namely d$_{yz}$-d$_{yz}$ and d$_{xy}$-d$_{xy}$ (cancelled), for J$_{1}$ and J$_{2}$ respectively}
    \label{FigS2: Electron Config Co$_{2}$+}
\end{figure}

\begin{figure}[ht]
    \centering
    \includegraphics[width=.7\linewidth]{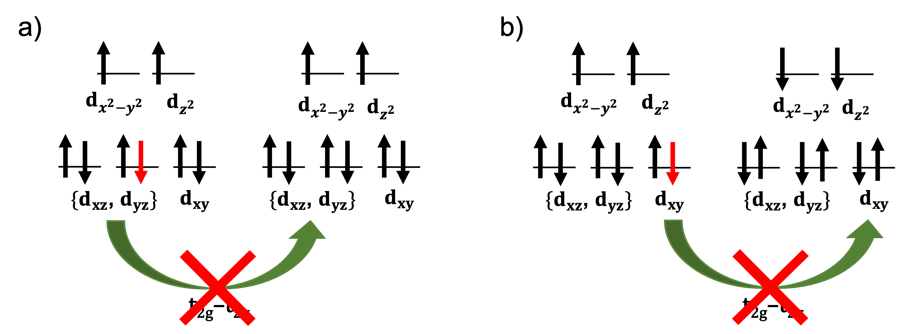}
    \caption{Electron configuration of the Ni$^{2+}$ magnetic ions connected by J$_{1}$ (a) and J$_{2}$ (b), showing parallel and antiparallel spin orientations, respectively. Green arrows illustrate the most relevant ferromagnetic superexchange channels, namely d$_{yz}$-d$_{yz}$ (cancelled) and d$_{xy}$-d$_{xy}$ (cancelled), for J$_{1}$ and J$_{2}$ respectively.}
    \label{FigS2: Electron Config Ni$_{2}$+}
\end{figure}

\begin{table}[]
\caption{CoPS$_{3}$, FePS$_{3}$ and NiPS$_{3}$ lattice parameters of the crystallographic non-magnetic (NM) and Stripy antiferromagnetic (AF) configurations}
\begin{tabular}{lcccccc}
                               & \multicolumn{2}{c}{\textbf{CoPS$_{3}$}} & \multicolumn{2}{c}{\textbf{FePS$_{3}$}} & \multicolumn{2}{c}{\textbf{NiPS$_{3}$}} \\ \hline
\textbf{Lattice parameter (\AA)} & \textbf{a}       & \textbf{b}      & \textbf{a}       & \textbf{b}      & \textbf{a}       & \textbf{b}      \\
\textbf{Crystallographic NM}   & 5.895            & 10.190          & 5.947            & 10.301          & 5.812            & 10.070          \\
\textbf{AF-stripy}            & 5.772            & 10.003          & 5.978            & 10.172          & 5.826            & 10.084          \\ \hline
\textbf{Change (\%)}           & -2.086           & -1.835          & +0.521           & -1.252          & +0.241           & +0.139          
\end{tabular}
\label{SItable1}
\end{table}

\begin{table}[H]
    \centering
    \caption{CoPS$_{3}$, FePS$_{3}$ and NiPS$_{3}$ lattice parameters of the crystallographic non-magnetic (NM) and Néel antiferromagnetic (AF) configurations.}
    \begin{tabular}{lcccccc}
    & \multicolumn{2}{c}{\textbf{CoPS$_{3}$}} & \multicolumn{2}{c}{\textbf{FePS$_{3}$}} & \multicolumn{2}{c}{\textbf{NiPS$_{3}$}} \\ \hline
    \textbf{Lattice parameter (\AA)} & \textbf{a}       & \textbf{b}      & \textbf{a}       & \textbf{b}      & \textbf{a}       & \textbf{b}      \\
    \textbf{Crystallographic NM}   & 5.895            & 10.190          & 5.947            & 10.301          & 5.812            & 10.070          \\
    \textbf{AF-N\'{e}el}            & 5.885            & 10.187          & 5.947            & 10.303          & 5.834            & 10.106          \\ \hline
    \textbf{Change (\%)}           & -0.169           & -0.029          & +0.000           & +0.019          & +0.378           & +0.357         
    \end{tabular}
    \label{SItable2}
\end{table}

\begin{table}[H]
    \centering
    \caption{CoPS$_{3}$, FePS$_{3}$ and NiPS$_{3}$ lattice parameters of the crystallographic non-magnetic (NM) and ferromagnetic (FM) configurations.}
    \begin{tabular}{lcccccc}
    & \multicolumn{2}{c}{\textbf{CoPS$_{3}$}} & \multicolumn{2}{c}{\textbf{FePS$_{3}$}} & \multicolumn{2}{c}{\textbf{NiPS$_{3}$}} \\ \hline
    \textbf{Lattice parameter (\AA)} & \textbf{a}       & \textbf{b}      & \textbf{a}       & \textbf{b}      & \textbf{a}       & \textbf{b}      \\
    \textbf{Crystallographic NM}   & 5.895            & 10.190          & 5.947            & 10.301          & 5.812            & 10.070          \\
    \textbf{FM}            & 5.836            & 10.103          & 5.932            & 10.278          & 5.824            & 10.089          \\ \hline
    \textbf{Change (\%)}           & -1.001          & -0.854         & -0.252          & -0.223         & +0.207          & +0.189        
    \end{tabular}
    \label{SItable3}
\end{table}

\begin{table}[H]
    \centering
    \caption{CoPS$_{3}$, FePS$_{3}$ and NiPS$_{3}$ distances (d and d’) of the crystallographic non-magnetic (NM) and zigzag antiferromagnetic (AF) configurations.}
    \begin{tabular}{lcccccc}
    & \multicolumn{2}{c}{\textbf{CoPS$_{3}$}} & \multicolumn{2}{c}{\textbf{FePS$_{3}$}} & \multicolumn{2}{c}{\textbf{NiPS$_{3}$}} \\ \hline
    \textbf{Distance (\AA)} & \textbf{d}      & \textbf{d'}      & \textbf{d}      & \textbf{d'}      & \textbf{d}      & \textbf{d'}      \\
    \textbf{Crystallographic NM}   & 3.397           & 3.402            & 3.433           & 3.433            & 3.357           & 3.356            \\
    \textbf{AF-zigzag}            & 3.723           & 3.192            & 3.626           & 3.316            & 3.353           & 3.359            \\ \hline
    \textbf{Change (\%)}           & +9.596          & -6.172           & +5.621         & -3.408          & -0.119          & +0.089          
    \end{tabular}
    \label{SItable4*}
\end{table}

\begin{table}[H]
    \centering
    \caption{CoPS$_{3}$, FePS$_{3}$ and NiPS$_{3}$ distances (d and d’) of the crystallographic non-magnetic (NM) and Stripy antiferromagnetic (AF) configurations.}
    \begin{tabular}{lcccccc}
    & \multicolumn{2}{c}{\textbf{CoPS$_{3}$}} & \multicolumn{2}{c}{\textbf{FePS$_{3}$}} & \multicolumn{2}{c}{\textbf{NiPS$_{3}$}} \\ \hline
    \textbf{Distance (\AA)} & \textbf{d}      & \textbf{d'}      & \textbf{d}      & \textbf{d'}      & \textbf{d}      & \textbf{d'}      \\
    \textbf{Crystallographic NM}   & 3.397           & 3.402            & 3.433           & 3.433            & 3.357           & 3.356            \\
    \textbf{AF-stripy}            & 3.337           & 3.332            & 2.670           & 3.843            & 3.373           & 3.357            \\ \hline
    \textbf{Change (\%)}           & -1.766          & -2.058           & -22.225         & +11.943          & +0.477          & +0.029          
    \end{tabular}
    \label{SItable4}
\end{table}

\begin{table}[H]
    \centering
    \caption{CoPS$_{3}$, FePS$_{3}$ and NiPS$_{3}$ distances (d and d’) of the crystallographic non-magnetic (NM) and Néel antiferromagnetic (AF) configurations.}
    \begin{tabular}{lcccccc}
    & \multicolumn{2}{c}{\textbf{CoPS$_{3}$}} & \multicolumn{2}{c}{\textbf{FePS$_{3}$}} & \multicolumn{2}{c}{\textbf{NiPS$_{3}$}} \\ \hline
    \textbf{Distance (\AA)} & \textbf{d}      & \textbf{d'}      & \textbf{d}      & \textbf{d'}      & \textbf{d}      & \textbf{d'}      \\
    \textbf{Crystallographic NM}   & 3.397           & 3.402            & 3.433           & 3.433            & 3.357           & 3.356            \\
    \textbf{AF-N\'{e}el}            & 3.393           & 3.398            & 3.436           & 3.433            & 3.369           & 3.369            \\ \hline
    \textbf{Change (\%)}           & -0.118          & -0.118           & +0.087          & +0.000           & +0.357          & +0.387          
    \end{tabular}
    \label{SItable5}
\end{table}

\begin{table}[H]
    \centering
    \caption{CoPS$_{3}$, FePS$_{3}$ and NiPS$_{3}$ distances (d and d’) of the crystallographic non-magnetic (NM) and ferromagnetic (FM) configurations.}
    \begin{tabular}{lcccccc}
    & \multicolumn{2}{c}{\textbf{CoPS$_{3}$}} & \multicolumn{2}{c}{\textbf{FePS$_{3}$}} & \multicolumn{2}{c}{\textbf{NiPS$_{3}$}} \\ \hline
    \textbf{Distance (\AA)} & \textbf{d}      & \textbf{d'}      & \textbf{d}      & \textbf{d'}      & \textbf{d}      & \textbf{d'}      \\
    \textbf{Crystallographic NM}   & 3.397           & 3.402            & 3.433           & 3.433            & 3.357           & 3.356            \\
    \textbf{FM}            & 3.368           & 3.368            & 3.429           & 3.424            & 3.363           & 3.362            \\ \hline
    \textbf{Change (\%)}           & -0.854          & -0.999           & +0.116          & +0.262           & +0.179          & +0.179          
    \end{tabular}
    \label{SItable6}
\end{table}

\begin{table}[H]
    \centering
    \caption{Orbital resolved exchange parameters for the experimental structure of FePS$_{3}$.}
    \begin{tabular}{c|rrrrrrrrrr}
\textbf{J$_{1}$ (meV)}    & \multicolumn{2}{r}{\textbf{d$_{z2}$}} & \multicolumn{2}{r}{\textbf{d$_{xz}$}} & \multicolumn{2}{r}{\textbf{d$_{yz}$}} & \multicolumn{2}{r}{\textbf{d$_{x2y2}$}} & \multicolumn{2}{r}{\textbf{d$_{xy}$}} \\ \hline
\textbf{d$_{z2}$}   & \multicolumn{2}{r}{-0.467}       & \multicolumn{2}{r}{-0.154}       & \multicolumn{2}{r}{-0.157}       & \multicolumn{2}{r}{-0.047}         & \multicolumn{2}{r}{-0.027}       \\
\textbf{d$_{xz}$}   & \multicolumn{2}{r}{-0.154}       & \multicolumn{2}{r}{7.091}        & \multicolumn{2}{r}{0.063}        & \multicolumn{2}{r}{-0.307}         & \multicolumn{2}{r}{0.288}        \\
\textbf{d$_{yz}$}   & \multicolumn{2}{r}{-0.157}       & \multicolumn{2}{r}{0.063}        & \multicolumn{2}{r}{7.215}        & \multicolumn{2}{r}{-0.103}         & \multicolumn{2}{r}{0.109}        \\
\textbf{d$_{x2y2}$} & \multicolumn{2}{r}{-0.047}       & \multicolumn{2}{r}{-0.307}       & \multicolumn{2}{r}{-0.103}       & \multicolumn{2}{r}{-0.194}         & \multicolumn{2}{r}{-0.017}       \\
\textbf{d$_{xy}$}   & \multicolumn{2}{r}{-0.027}       & \multicolumn{2}{r}{0.288}        & \multicolumn{2}{r}{0.109}        & \multicolumn{2}{r}{-0.017}         & \multicolumn{2}{r}{1.612}       
\end{tabular}
    \label{SItable7}
\end{table}

\begin{table}[H]
    \centering
    \caption{Orbital resolved exchange parameters for the experimental structure of FePS$_{3}$.}
    \begin{tabular}{c|rrrrrrrrrr}
    \textbf{J$_{2}$ (meV)}            & \multicolumn{2}{r}{\textbf{d$_{z2}$}} & \multicolumn{2}{r}{\textbf{d$_{xz}$}} & \multicolumn{2}{r}{\textbf{d$_{yz}$}} & \multicolumn{2}{r}{\textbf{d$_{x2y2}$}} & \multicolumn{2}{r}{\textbf{d$_{xy}$}} \\ \hline
    \textbf{d$_{z2}$}   & \multicolumn{2}{r}{-0.045}       & \multicolumn{2}{r}{-0.013}       & \multicolumn{2}{r}{-0.011}       & \multicolumn{2}{r}{-0.008}         & \multicolumn{2}{r}{-1.18}        \\
    \textbf{d$_{xz}$}   & \multicolumn{2}{r}{-0.013}       & \multicolumn{2}{r}{0.334}        & \multicolumn{2}{r}{0.14}         & \multicolumn{2}{r}{0.0}            & \multicolumn{2}{r}{0.791}        \\
    \textbf{d$_{yz}$}   & \multicolumn{2}{r}{-0.011}       & \multicolumn{2}{r}{0.14}         & \multicolumn{2}{r}{0.418}        & \multicolumn{2}{r}{0.036}          & \multicolumn{2}{r}{0.634}        \\
    \textbf{d$_{x2y2}$} & \multicolumn{2}{r}{-0.008}       & \multicolumn{2}{r}{0.0}          & \multicolumn{2}{r}{0.036}        & \multicolumn{2}{r}{-0.783}         & \multicolumn{2}{r}{0.086}        \\
    \textbf{d$_{xy}$}   & \multicolumn{2}{r}{-1.18}        & \multicolumn{2}{r}{0.791}        & \multicolumn{2}{r}{0.634}        & \multicolumn{2}{r}{0.086}          & \multicolumn{2}{r}{11.016}      
    \end{tabular}
    \label{SItable8}
\end{table}

\begin{table}[H]
    \centering
    \caption{Orbital resolved exchange parameters for the experimental structure of FePS$_{3}$.}
    \begin{tabular}{c|rrrrrrrrrr}
\textbf{J$_{3}$ (meV)}    & \multicolumn{2}{r}{\textbf{d$_{z2}$}} & \multicolumn{2}{r}{\textbf{d$_{xz}$}} & \multicolumn{2}{r}{\textbf{d$_{yz}$}} & \multicolumn{2}{r}{\textbf{d$_{x2y2}$}} & \multicolumn{2}{r}{\textbf{d$_{xy}$}} \\ \hline
\textbf{d$_{z2}$}   & \multicolumn{2}{r}{0.01}         & \multicolumn{2}{r}{-0.058}       & \multicolumn{2}{r}{-0.065}       & \multicolumn{2}{r}{-0.028}         & \multicolumn{2}{r}{-0.011}       \\
\textbf{d$_{xz}$}   & \multicolumn{2}{r}{-0.058}       & \multicolumn{2}{r}{-0.02}        & \multicolumn{2}{r}{-0.037}       & \multicolumn{2}{r}{-0.101}         & \multicolumn{2}{r}{-0.076}       \\
\textbf{d$_{yz}$}   & \multicolumn{2}{r}{-0.065}       & \multicolumn{2}{r}{-0.037}       & \multicolumn{2}{r}{-0.022}       & \multicolumn{2}{r}{-0.122}         & \multicolumn{2}{r}{0.089}        \\
\textbf{d$_{x2y2}$} & \multicolumn{2}{r}{-0.028}       & \multicolumn{2}{r}{-0.101}       & \multicolumn{2}{r}{-0.122}       & \multicolumn{2}{r}{-0.004}         & \multicolumn{2}{r}{-0.041}       \\
\textbf{d$_{xy}$}   & \multicolumn{2}{r}{-0.011}       & \multicolumn{2}{r}{-0.076}       & \multicolumn{2}{r}{-0.089}       & \multicolumn{2}{r}{-0.041}         & \multicolumn{2}{r}{0.405}       
\end{tabular}
    \label{SItable9}
\end{table}

\begin{table}[H]
    \centering
    \caption{Orbital resolved exchange parameters for the experimental structure of FePS$_{3}$.}
    \begin{tabular}{c|rrrrrrrrrr}
\textbf{J$_{4}$ (meV)}    & \multicolumn{2}{r}{\textbf{d$_{z2}$}} & \multicolumn{2}{r}{\textbf{d$_{xz}$}} & \multicolumn{2}{r}{\textbf{d$_{yz}$}} & \multicolumn{2}{r}{\textbf{d$_{x2y2}$}} & \multicolumn{2}{r}{\textbf{d$_{xy}$}} \\ \hline
\textbf{d$_{z2}$}   & \multicolumn{2}{r}{-0.121}       & \multicolumn{2}{r}{0.002}        & \multicolumn{2}{r}{0.003}        & \multicolumn{2}{r}{0.013}          & \multicolumn{2}{r}{-0.234}       \\
\textbf{d$_{xz}$}   & \multicolumn{2}{r}{0.002}        & \multicolumn{2}{r}{-0.093}       & \multicolumn{2}{r}{-0.022}       & \multicolumn{2}{r}{-0.005}         & \multicolumn{2}{r}{-0.146}       \\
\textbf{d$_{yz}$}   & \multicolumn{2}{r}{0.003}        & \multicolumn{2}{r}{-0.022}       & \multicolumn{2}{r}{-0.081}       & \multicolumn{2}{r}{0.002}          & \multicolumn{2}{r}{0.122}        \\
\textbf{d$_{x2y2}$} & \multicolumn{2}{r}{0.012}        & \multicolumn{2}{r}{-0.005}       & \multicolumn{2}{r}{0.002}        & \multicolumn{2}{r}{-0.01}          & \multicolumn{2}{r}{-0.18}        \\
\textbf{d$_{xy}$}   & \multicolumn{2}{r}{-0.234}       & \multicolumn{2}{r}{-0.146}       & \multicolumn{2}{r}{-0.122}       & \multicolumn{2}{r}{-0.18}          & \multicolumn{2}{r}{-2.509}      
\end{tabular}
    \label{SItable10}
\end{table}

\begin{table}[H]
    \centering
    \caption{Orbital resolved exchange parameters for the experimental structure of FePS$_{3}$.}
    \begin{tabular}{c|rrrrrrrrrr}
\textbf{J$_{5}$ (meV)}    & \multicolumn{2}{r}{\textbf{d$_{z2}$}} & \multicolumn{2}{r}{\textbf{d$_{xz}$}} & \multicolumn{2}{r}{\textbf{d$_{yz}$}} & \multicolumn{2}{r}{\textbf{d$_{x2y2}$}} & \multicolumn{2}{r}{\textbf{d$_{xy}$}} \\ \hline
\textbf{d$_{z2}$}   & \multicolumn{2}{r}{-2.323}       & \multicolumn{2}{r}{0.019}        & \multicolumn{2}{r}{0.015}        & \multicolumn{2}{r}{-1.175}         & \multicolumn{2}{r}{0.148}        \\
\textbf{d$_{xz}$}   & \multicolumn{2}{r}{0.019}        & \multicolumn{2}{r}{0.035}        & \multicolumn{2}{r}{0.024}        & \multicolumn{2}{r}{-0.01}          & \multicolumn{2}{r}{-0.239}       \\
\textbf{d$_{yz}$}   & \multicolumn{2}{r}{0.015}        & \multicolumn{2}{r}{0.024}        & \multicolumn{2}{r}{0.022}        & \multicolumn{2}{r}{-0.032}         & \multicolumn{2}{r}{0.154}        \\
\textbf{d$_{x2y2}$} & \multicolumn{2}{r}{-1.175}       & \multicolumn{2}{r}{-0.01}        & \multicolumn{2}{r}{-0.032}       & \multicolumn{2}{r}{-0.154}         & \multicolumn{2}{r}{-0.138}       \\
\textbf{d$_{xy}$}   & \multicolumn{2}{r}{0.148}        & \multicolumn{2}{r}{-0.239}       & \multicolumn{2}{r}{-0.154}       & \multicolumn{2}{r}{-0.138}         & \multicolumn{2}{r}{-2.491}      
\end{tabular}
    \label{SItable11}
\end{table}

\begin{table}[H]
    \centering
    \caption{Orbital resolved exchange parameters for the experimental structure of FePS$_{3}$.}
    \begin{tabular}{c|rrrrrrrrrr}
\textbf{J$_{6}$ (meV)}    & \multicolumn{2}{r}{\textbf{d$_{z2}$}} & \multicolumn{2}{r}{\textbf{d$_{xz}$}} & \multicolumn{2}{r}{\textbf{d$_{yz}$}} & \multicolumn{2}{r}{\textbf{d$_{x2y2}$}} & \multicolumn{2}{r}{\textbf{d$_{xy}$}} \\ \hline
\textbf{d$_{z2}$}   & \multicolumn{2}{r}{-0.156}       & \multicolumn{2}{r}{0.03}         & \multicolumn{2}{r}{0.025}        & \multicolumn{2}{r}{-0.0}           & \multicolumn{2}{r}{-0.05}        \\
\textbf{d$_{xz}$}   & \multicolumn{2}{r}{0.03}         & \multicolumn{2}{r}{-0.012}       & \multicolumn{2}{r}{0.004}        & \multicolumn{2}{r}{0.053}          & \multicolumn{2}{r}{-0.073}       \\
\textbf{d$_{yz}$}   & \multicolumn{2}{r}{0.025}        & \multicolumn{2}{r}{0.004}        & \multicolumn{2}{r}{-0.01}        & \multicolumn{2}{r}{0.04}           & \multicolumn{2}{r}{0.042}        \\
\textbf{d$_{x2y2}$} & \multicolumn{2}{r}{-0.0}         & \multicolumn{2}{r}{0.053}        & \multicolumn{2}{r}{0.04}         & \multicolumn{2}{r}{-5.914}         & \multicolumn{2}{r}{-0.044}       \\
\textbf{d$_{xy}$}   & \multicolumn{2}{r}{-0.05}        & \multicolumn{2}{r}{-0.073}       & \multicolumn{2}{r}{-0.042}       & \multicolumn{2}{r}{-0.044}         & \multicolumn{2}{r}{0.325}       
\end{tabular}
    \label{SItable12}
\end{table}

\begin{table}[H]
    \centering
    \caption{Orbital resolved exchange parameters for the spin polarized optimized structure of FePS$_{3}$.}
    \begin{tabular}{c|rrrrrrrrrr}
\textbf{J$_{1}$ (meV)}    & \multicolumn{2}{r}{\textbf{d$_{z2}$}} & \multicolumn{2}{r}{\textbf{d$_{xz}$}} & \multicolumn{2}{r}{\textbf{d$_{yz}$}} & \multicolumn{2}{r}{\textbf{d$_{x2y2}$}} & \multicolumn{2}{r}{\textbf{d$_{xy}$}} \\ \hline
\textbf{d$_{z2}$}   & \multicolumn{2}{r}{-0.899}       & \multicolumn{2}{r}{-0.192}       & \multicolumn{2}{r}{-0.188}       & \multicolumn{2}{r}{-0.127}         & \multicolumn{2}{r}{-0.017}       \\
\textbf{d$_{xz}$}   & \multicolumn{2}{r}{-0.192}       & \multicolumn{2}{r}{8.323}        & \multicolumn{2}{r}{0.035}        & \multicolumn{2}{r}{-0.29}          & \multicolumn{2}{r}{0.306}        \\
\textbf{d$_{yz}$}   & \multicolumn{2}{r}{-0.188}       & \multicolumn{2}{r}{0.035}        & \multicolumn{2}{r}{8.383}        & \multicolumn{2}{r}{-0.088}         & \multicolumn{2}{r}{0.12}         \\
\textbf{d$_{x2y2}$} & \multicolumn{2}{r}{-0.127}       & \multicolumn{2}{r}{-0.29}        & \multicolumn{2}{r}{-0.088}       & \multicolumn{2}{r}{-0.28}          & \multicolumn{2}{r}{-0.044}       \\
\textbf{d$_{xy}$}   & \multicolumn{2}{r}{-0.017}       & \multicolumn{2}{r}{0.306}        & \multicolumn{2}{r}{0.12}         & \multicolumn{2}{r}{-0.044}         & \multicolumn{2}{r}{1.823}       
\end{tabular}
    \label{SItable13}
\end{table}

\begin{table}[H]
    \centering
    \caption{Orbital resolved exchange parameters for the spin polarized optimized structure of FePS$_{3}$.}
    \begin{tabular}{c|rrrrrrrrrr}
\textbf{J$_{2}$ (meV)}    & \multicolumn{2}{r}{\textbf{d$_{z2}$}} & \multicolumn{2}{r}{\textbf{d$_{xz}$}} & \multicolumn{2}{r}{\textbf{d$_{yz}$}} & \multicolumn{2}{r}{\textbf{d$_{x2y2}$}} & \multicolumn{2}{r}{\textbf{d$_{xy}$}} \\ \hline
\textbf{d$_{z2}$}   & \multicolumn{2}{r}{-0.104}       & \multicolumn{2}{r}{-0.021}       & \multicolumn{2}{r}{-0.019}       & \multicolumn{2}{r}{-0.009}         & \multicolumn{2}{r}{-2.114}       \\
\textbf{d$_{xz}$}   & \multicolumn{2}{r}{-0.021}       & \multicolumn{2}{r}{0.161}        & \multicolumn{2}{r}{0.019}        & \multicolumn{2}{r}{-0.005}         & \multicolumn{2}{r}{0.5}          \\
\textbf{d$_{yz}$}   & \multicolumn{2}{r}{-0.019}       & \multicolumn{2}{r}{0.019}        & \multicolumn{2}{r}{0.182}        & \multicolumn{2}{r}{0.002}          & \multicolumn{2}{r}{0.432}        \\
\textbf{d$_{x2y2}$} & \multicolumn{2}{r}{-0.009}       & \multicolumn{2}{r}{-0.005}       & \multicolumn{2}{r}{0.002}        & \multicolumn{2}{r}{0.128}          & \multicolumn{2}{r}{0.015}        \\
\textbf{d$_{xy}$}   & \multicolumn{2}{r}{-2.114}       & \multicolumn{2}{r}{0.5}          & \multicolumn{2}{r}{0.432}        & \multicolumn{2}{r}{0.015}          & \multicolumn{2}{r}{5.022}       
\end{tabular}
    \label{SItable14}
\end{table}

\begin{table}[H]
    \centering
    \caption{Orbital resolved exchange parameters for the spin polarized optimized structure of FePS$_{3}$.}
    \begin{tabular}{c|rlrlrlrlrl}
\textbf{J$_{3}$ (meV)}    & \multicolumn{2}{r}{\textbf{d$_{z2}$}} & \multicolumn{2}{r}{\textbf{d$_{xz}$}} & \multicolumn{2}{r}{\textbf{d$_{yz}$}} & \multicolumn{2}{r}{\textbf{d$_{x2y2}$}} & \multicolumn{2}{r}{\textbf{d$_{xy}$}} \\ \hline
\textbf{d$_{z2}$}   & \multicolumn{2}{r}{0.012}        & \multicolumn{2}{r}{-0.083}       & \multicolumn{2}{r}{-0.093}       & \multicolumn{2}{r}{-0.059}         & \multicolumn{2}{r}{-0.013}       \\
\textbf{d$_{xz}$}   & \multicolumn{2}{r}{-0.083}       & \multicolumn{2}{r}{-0.025}       & \multicolumn{2}{r}{-0.044}       & \multicolumn{2}{r}{-0.088}         & \multicolumn{2}{r}{-0.059}       \\
\textbf{d$_{yz}$}   & \multicolumn{2}{r}{-0.093}       & \multicolumn{2}{r}{-0.044}       & \multicolumn{2}{r}{-0.027}       & \multicolumn{2}{r}{-0.105}         & \multicolumn{2}{r}{-0.064}       \\
\textbf{d$_{x2y2}$} & \multicolumn{2}{r}{-0.059}       & \multicolumn{2}{r}{-0.088}       & \multicolumn{2}{r}{-0.105}       & \multicolumn{2}{r}{0.007}          & \multicolumn{2}{r}{-0.038}       \\
\textbf{d$_{xy}$}   & \multicolumn{2}{r}{-0.013}       & \multicolumn{2}{r}{-0.059}       & \multicolumn{2}{r}{-0.064}       & \multicolumn{2}{r}{-0.038}         & \multicolumn{2}{r}{0.278}       
\end{tabular}
    \label{SItable15}
\end{table}

\begin{table}[H]
    \centering
    \caption{Orbital resolved exchange parameters for the spin polarized optimized structure of FePS$_{3}$.}
    \begin{tabular}{c|rlrlrlrlrl}
\textbf{J$_{4}$ (meV)}    & \multicolumn{2}{r}{\textbf{d$_{z2}$}} & \multicolumn{2}{r}{\textbf{d$_{xz}$}} & \multicolumn{2}{r}{\textbf{d$_{yz}$}} & \multicolumn{2}{r}{\textbf{d$_{x2y2}$}} & \multicolumn{2}{r}{\textbf{d$_{xy}$}} \\ \hline
\textbf{d$_{z2}$}   & \multicolumn{2}{r}{-0.11}        & \multicolumn{2}{r}{0.002}        & \multicolumn{2}{r}{0.003}        & \multicolumn{2}{r}{0.036}          & \multicolumn{2}{r}{-0.212}       \\
\textbf{d$_{xz}$}   & \multicolumn{2}{r}{0.002}        & \multicolumn{2}{r}{-0.112}       & \multicolumn{2}{r}{-0.039}       & \multicolumn{2}{r}{-0.014}         & \multicolumn{2}{r}{-0.061}       \\
\textbf{d$_{yz}$}   & \multicolumn{2}{r}{0.003}        & \multicolumn{2}{r}{-0.039}       & \multicolumn{2}{r}{-0.105}       & \multicolumn{2}{r}{0.0}            & \multicolumn{2}{r}{-0.066}       \\
\textbf{d$_{x2y2}$} & \multicolumn{2}{r}{0.036}        & \multicolumn{2}{r}{-0.014}       & \multicolumn{2}{r}{0.0}          & \multicolumn{2}{r}{-0.004}         & \multicolumn{2}{r}{-0.164}       \\
\textbf{d$_{xy}$}   & \multicolumn{2}{r}{-0.212}       & \multicolumn{2}{r}{-0.061}       & \multicolumn{2}{r}{-0.066}       & \multicolumn{2}{r}{-0.164}         & \multicolumn{2}{r}{-1.186}      
\end{tabular}
    \label{SItable16}
\end{table}

\begin{table}[H]
    \centering
    \caption{Orbital resolved exchange parameters for the spin polarized optimized structure of FePS$_{3}$.}
    \begin{tabular}{c|rlrlrlrlrl}
\textbf{J$_{5}$ (meV)}    & \multicolumn{2}{r}{\textbf{d$_{z2}$}} & \multicolumn{2}{r}{\textbf{d$_{xz}$}} & \multicolumn{2}{r}{\textbf{d$_{yz}$}} & \multicolumn{2}{r}{\textbf{d$_{x2y2}$}} & \multicolumn{2}{r}{\textbf{d$_{xy}$}} \\ \hline
\textbf{d$_{z2}$}   & \multicolumn{2}{r}{-2.613}       & \multicolumn{2}{r}{-0.016}       & \multicolumn{2}{r}{-0.014}       & \multicolumn{2}{r}{-1.223}         & \multicolumn{2}{r}{0.154}        \\
\textbf{d$_{xz}$}   & \multicolumn{2}{r}{-0.016}       & \multicolumn{2}{r}{-0.008}       & \multicolumn{2}{r}{0.032}        & \multicolumn{2}{r}{-0.059}         & \multicolumn{2}{r}{-0.105}       \\
\textbf{d$_{yz}$}   & \multicolumn{2}{r}{-0.014}       & \multicolumn{2}{r}{0.032}        & \multicolumn{2}{r}{-0.016}       & \multicolumn{2}{r}{-0.075}         & \multicolumn{2}{r}{-0.06}        \\
\textbf{d$_{x2y2}$} & \multicolumn{2}{r}{-1.223}       & \multicolumn{2}{r}{-0.059}       & \multicolumn{2}{r}{-0.075}       & \multicolumn{2}{r}{-0.181}         & \multicolumn{2}{r}{-0.122}       \\
\textbf{d$_{xy}$}   & \multicolumn{2}{r}{0.154}        & \multicolumn{2}{r}{-0.105}       & \multicolumn{2}{r}{-0.06}        & \multicolumn{2}{r}{-0.122}         & \multicolumn{2}{r}{-1.114}      
\end{tabular}
    \label{SItable17}
\end{table}

\begin{table}[H]
    \centering
    \caption{Orbital resolved exchange parameters for the spin polarized optimized structure of FePS$_{3}$.}
    \begin{tabular}{c|rlrlrlrlrl}
\textbf{J$_{6}$ (meV)}    & \multicolumn{2}{r}{\textbf{d$_{z2}$}} & \multicolumn{2}{r}{\textbf{d$_{xz}$}} & \multicolumn{2}{r}{\textbf{d$_{yz}$}} & \multicolumn{2}{r}{\textbf{d$_{x2y2}$}} & \multicolumn{2}{r}{\textbf{d$_{xy}$}} \\ \hline
\textbf{d$_{z2}$}   & \multicolumn{2}{r}{-0.196}       & \multicolumn{2}{r}{0.064}        & \multicolumn{2}{r}{0.056}        & \multicolumn{2}{r}{-0.001}         & \multicolumn{2}{r}{-0.236}       \\
\textbf{d$_{xz}$}   & \multicolumn{2}{r}{0.064}        & \multicolumn{2}{r}{-0.026}       & \multicolumn{2}{r}{0.005}        & \multicolumn{2}{r}{0.083}          & \multicolumn{2}{r}{-0.092}       \\
\textbf{d$_{yz}$}   & \multicolumn{2}{r}{0.056}        & \multicolumn{2}{r}{0.005}        & \multicolumn{2}{r}{-0.022}       & \multicolumn{2}{r}{0.083}          & \multicolumn{2}{r}{-0.071}       \\
\textbf{d$_{x2y2}$} & \multicolumn{2}{r}{-0.001}       & \multicolumn{2}{r}{0.083}        & \multicolumn{2}{r}{0.083}        & \multicolumn{2}{r}{-5.386}         & \multicolumn{2}{r}{-0.025}       \\
\textbf{d$_{xy}$}   & \multicolumn{2}{r}{-0.236}       & \multicolumn{2}{r}{-0.092}       & \multicolumn{2}{r}{-0.071}       & \multicolumn{2}{r}{-0.025}         & \multicolumn{2}{r}{1.036}       
\end{tabular}
    \label{SItable18}
\end{table}

\begin{table}[H]
    \centering
    \caption{Orbital resolved exchange parameters for the experimental structure of CoPS$_{3}$.}
    \begin{tabular}{c|rlrlrlrlrl}
\textbf{J$_{1}$ (meV)}    & \multicolumn{2}{r}{\textbf{d$_{z2}$}} & \multicolumn{2}{r}{\textbf{d$_{xz}$}} & \multicolumn{2}{r}{\textbf{d$_{yz}$}} & \multicolumn{2}{r}{\textbf{d$_{x2y2}$}} & \multicolumn{2}{r}{\textbf{d$_{xy}$}} \\ \hline
\textbf{d$_{z2}$}   & \multicolumn{2}{r}{0.168}        & \multicolumn{2}{r}{-0.907}       & \multicolumn{2}{r}{-0.922}       & \multicolumn{2}{r}{0.108}          & \multicolumn{2}{r}{-0.221}       \\
\textbf{d$_{xz}$}   & \multicolumn{2}{r}{-0.907}       & \multicolumn{2}{r}{7.514}        & \multicolumn{2}{r}{0.105}        & \multicolumn{2}{r}{-0.977}         & \multicolumn{2}{r}{-0.417}       \\
\textbf{d$_{yz}$}   & \multicolumn{2}{r}{-0.922}       & \multicolumn{2}{r}{0.105}        & \multicolumn{2}{r}{7.465}        & \multicolumn{2}{r}{-1.266}         & \multicolumn{2}{r}{-0.066}       \\
\textbf{d$_{x2y2}$} & \multicolumn{2}{r}{0.108}        & \multicolumn{2}{r}{-0.977}       & \multicolumn{2}{r}{-1.266}       & \multicolumn{2}{r}{-0.094}         & \multicolumn{2}{r}{0.073}        \\
\textbf{d$_{xy}$}   & \multicolumn{2}{r}{-0.221}       & \multicolumn{2}{r}{-0.417}       & \multicolumn{2}{r}{-0.065}       & \multicolumn{2}{r}{0.073}          & \multicolumn{2}{r}{-0.36}       
\end{tabular}
    \label{SItable19}
\end{table}

\begin{table}[H]
    \centering
    \caption{Orbital resolved exchange parameters for the experimental structure of CoPS$_{3}$.}
    \begin{tabular}{c|rlrlrlrlrl}
\textbf{J$_{2}$ (meV)}    & \multicolumn{2}{r}{\textbf{d$_{z2}$}} & \multicolumn{2}{r}{\textbf{d$_{xz}$}} & \multicolumn{2}{r}{\textbf{d$_{yz}$}} & \multicolumn{2}{r}{\textbf{d$_{x2y2}$}} & \multicolumn{2}{r}{\textbf{d$_{xy}$}} \\ \hline
\textbf{d$_{z2}$}   & \multicolumn{2}{r}{-0.371}       & \multicolumn{2}{r}{0.078}        & \multicolumn{2}{r}{0.066}        & \multicolumn{2}{r}{-0.002}         & \multicolumn{2}{r}{-0.242}       \\
\textbf{d$_{xz}$}   & \multicolumn{2}{r}{0.078}        & \multicolumn{2}{r}{0.524}        & \multicolumn{2}{r}{0.022}        & \multicolumn{2}{r}{0.115}          & \multicolumn{2}{r}{-0.093}       \\
\textbf{d$_{yz}$}   & \multicolumn{2}{r}{0.066}        & \multicolumn{2}{r}{0.022}        & \multicolumn{2}{r}{0.54}         & \multicolumn{2}{r}{-0.106}         & \multicolumn{2}{r}{0.126}        \\
\textbf{d$_{x2y2}$} & \multicolumn{2}{r}{-0.002}       & \multicolumn{2}{r}{0.115}        & \multicolumn{2}{r}{-0.106}       & \multicolumn{2}{r}{-0.687}         & \multicolumn{2}{r}{-0.034}       \\
\textbf{d$_{xy}$}   & \multicolumn{2}{r}{-0.242}       & \multicolumn{2}{r}{-0.093}       & \multicolumn{2}{r}{0.126}        & \multicolumn{2}{r}{-0.034}         & \multicolumn{2}{r}{-0.068}      
\end{tabular}
    
    \label{SItable20}
\end{table}

\begin{table}[H]
    \centering
    \caption{Orbital resolved exchange parameters for the experimental structure of CoPS$_{3}$.}
    \begin{tabular}{c|rlrlrlrlrl}
\textbf{J$_{3}$ (meV)}    & \multicolumn{2}{r}{\textbf{d$_{z2}$}} & \multicolumn{2}{r}{\textbf{d$_{xz}$}} & \multicolumn{2}{r}{\textbf{d$_{yz}$}} & \multicolumn{2}{r}{\textbf{d$_{x2y2}$}} & \multicolumn{2}{r}{\textbf{d$_{xy}$}} \\ \hline
\textbf{d$_{z2}$}   & \multicolumn{2}{r}{0.069}        & \multicolumn{2}{r}{-0.117}       & \multicolumn{2}{r}{-0.114}       & \multicolumn{2}{r}{-0.047}         & \multicolumn{2}{r}{-0.004}       \\
\textbf{d$_{xz}$}   & \multicolumn{2}{r}{-0.117}       & \multicolumn{2}{r}{-0.06}        & \multicolumn{2}{r}{-0.024}       & \multicolumn{2}{r}{-0.326}         & \multicolumn{2}{r}{-0.003}       \\
\textbf{d$_{yz}$}   & \multicolumn{2}{r}{-0.114}       & \multicolumn{2}{r}{-0.024}       & \multicolumn{2}{r}{-0.059}       & \multicolumn{2}{r}{-0.367}         & \multicolumn{2}{r}{0.01}         \\
\textbf{d$_{x2y2}$} & \multicolumn{2}{r}{-0.047}       & \multicolumn{2}{r}{-0.326}       & \multicolumn{2}{r}{-0.367}       & \multicolumn{2}{r}{-0.17}          & \multicolumn{2}{r}{0.003}        \\
\textbf{d$_{xy}$}   & \multicolumn{2}{r}{-0.004}       & \multicolumn{2}{r}{-0.003}       & \multicolumn{2}{r}{0.01}         & \multicolumn{2}{r}{0.004}          & \multicolumn{2}{r}{-0.022}      
\end{tabular}
    
    \label{SItable21}
\end{table}

\begin{table}[H]
    \centering
    \caption{Orbital resolved exchange parameters for the experimental structure of CoPS$_{3}$.}
    \begin{tabular}{c|rlrlrlrlrl}
\textbf{J$_{4}$ (meV)}    & \multicolumn{2}{r}{\textbf{d$_{z2}$}} & \multicolumn{2}{r}{\textbf{d$_{xz}$}} & \multicolumn{2}{r}{\textbf{d$_{yz}$}} & \multicolumn{2}{r}{\textbf{d$_{x2y2}$}} & \multicolumn{2}{r}{\textbf{d$_{xy}$}} \\ \hline
\textbf{d$_{z2}$}   & \multicolumn{2}{r}{-0.025}       & \multicolumn{2}{r}{-0.011}       & \multicolumn{2}{r}{-0.001}       & \multicolumn{2}{r}{0.015}          & \multicolumn{2}{r}{0.046}        \\
\textbf{d$_{xz}$}   & \multicolumn{2}{r}{-0.011}       & \multicolumn{2}{r}{-0.062}       & \multicolumn{2}{r}{-0.004}       & \multicolumn{2}{r}{0.002}          & \multicolumn{2}{r}{0.047}        \\
\textbf{d$_{yz}$}   & \multicolumn{2}{r}{-0.001}       & \multicolumn{2}{r}{-0.004}       & \multicolumn{2}{r}{-0.05}        & \multicolumn{2}{r}{-0.006}         & \multicolumn{2}{r}{0.046}        \\
\textbf{d$_{x2y2}$} & \multicolumn{2}{r}{0.015}        & \multicolumn{2}{r}{0.002}        & \multicolumn{2}{r}{-0.006}       & \multicolumn{2}{r}{-0.082}         & \multicolumn{2}{r}{-0.013}       \\
\textbf{d$_{xy}$}   & \multicolumn{2}{r}{0.046}        & \multicolumn{2}{r}{0.047}        & \multicolumn{2}{r}{0.046}        & \multicolumn{2}{r}{-0.013}         & \multicolumn{2}{r}{0.055}       
\end{tabular}
    
    \label{SItable22}
\end{table}

\begin{table}[H]
    \centering
    \caption{Orbital resolved exchange parameters for the experimental structure of CoPS$_{3}$.}
    \begin{tabular}{c|rlrlrlrlrl}
\textbf{J$_{5}$ (meV)}    & \multicolumn{2}{r}{\textbf{d$_{z2}$}} & \multicolumn{2}{r}{\textbf{d$_{xz}$}} & \multicolumn{2}{r}{\textbf{d$_{yz}$}} & \multicolumn{2}{r}{\textbf{d$_{x2y2}$}} & \multicolumn{2}{r}{\textbf{d$_{xy}$}} \\ \hline
\textbf{d$_{z2}$}   & \multicolumn{2}{r}{-3.482}       & \multicolumn{2}{r}{-0.181}       & \multicolumn{2}{r}{-0.163}       & \multicolumn{2}{r}{-0.862}         & \multicolumn{2}{r}{0.13}         \\
\textbf{d$_{xz}$}   & \multicolumn{2}{r}{-0.181}       & \multicolumn{2}{r}{-0.018}       & \multicolumn{2}{r}{0.001}        & \multicolumn{2}{r}{-0.263}         & \multicolumn{2}{r}{0.034}        \\
\textbf{d$_{yz}$}   & \multicolumn{2}{r}{-0.163}       & \multicolumn{2}{r}{0.001}        & \multicolumn{2}{r}{-0.044}       & \multicolumn{2}{r}{-0.18}          & \multicolumn{2}{r}{0.004}        \\
\textbf{d$_{x2y2}$} & \multicolumn{2}{r}{-0.862}       & \multicolumn{2}{r}{-0.263}       & \multicolumn{2}{r}{-0.18}        & \multicolumn{2}{r}{-0.016}         & \multicolumn{2}{r}{0.097}        \\
\textbf{d$_{xy}$}   & \multicolumn{2}{r}{0.13}         & \multicolumn{2}{r}{0.034}        & \multicolumn{2}{r}{0.004}        & \multicolumn{2}{r}{0.097}          & \multicolumn{2}{r}{0.051}       
\end{tabular}
    
    \label{SItable23}
\end{table}

\begin{table}[H]
    \centering
    \caption{Orbital resolved exchange parameters for the experimental structure of CoPS$_{3}$.}
    \begin{tabular}{c|rlrlrlrlrl}
\textbf{J$_{3}$ (meV)}    & \multicolumn{2}{r}{\textbf{d$_{z2}$}} & \multicolumn{2}{r}{\textbf{d$_{xz}$}} & \multicolumn{2}{r}{\textbf{d$_{yz}$}} & \multicolumn{2}{r}{\textbf{d$_{x2y2}$}} & \multicolumn{2}{r}{\textbf{d$_{xy}$}} \\ \hline
\textbf{d$_{z2}$}   & \multicolumn{2}{r}{-0.321}       & \multicolumn{2}{r}{0.054}        & \multicolumn{2}{r}{0.045}        & \multicolumn{2}{r}{0.001}          & \multicolumn{2}{r}{0.114}        \\
\textbf{d$_{xz}$}   & \multicolumn{2}{r}{0.054}        & \multicolumn{2}{r}{-0.018}       & \multicolumn{2}{r}{-0.023}       & \multicolumn{2}{r}{0.138}          & \multicolumn{2}{r}{0.072}        \\
\textbf{d$_{yz}$}   & \multicolumn{2}{r}{0.045}        & \multicolumn{2}{r}{-0.023}       & \multicolumn{2}{r}{-0.01}        & \multicolumn{2}{r}{0.224}          & \multicolumn{2}{r}{0.01}         \\
\textbf{d$_{x2y2}$} & \multicolumn{2}{r}{0.001}        & \multicolumn{2}{r}{0.138}        & \multicolumn{2}{r}{0.224}        & \multicolumn{2}{r}{-3.745}         & \multicolumn{2}{r}{-0.026}       \\
\textbf{d$_{xy}$}   & \multicolumn{2}{r}{0.114}        & \multicolumn{2}{r}{0.072}        & \multicolumn{2}{r}{0.01}         & \multicolumn{2}{r}{-0.026}         & \multicolumn{2}{r}{0.029}       
\end{tabular}
    
    \label{SItable24}
\end{table}

\begin{table}[H]
    \centering
    \caption{Orbital resolved exchange parameters for the spin polarized optimized structure of CoPS$_{3}$.}
    \begin{tabular}{c|rlrlrlrlrl}
\textbf{J$_{1}$ (meV)}    & \multicolumn{2}{r}{\textbf{d$_{z2}$}} & \multicolumn{2}{r}{\textbf{d$_{xz}$}} & \multicolumn{2}{r}{\textbf{d$_{yz}$}} & \multicolumn{2}{r}{\textbf{d$_{x2y2}$}} & \multicolumn{2}{r}{\textbf{d$_{xy}$}} \\ \hline
\textbf{d$_{z2}$}   & \multicolumn{2}{r}{-0.029}       & \multicolumn{2}{r}{-0.889}       & \multicolumn{2}{r}{-0.868}       & \multicolumn{2}{r}{-0.449}         & \multicolumn{2}{r}{-0.182}       \\
\textbf{d$_{xz}$}   & \multicolumn{2}{r}{-0.889}       & \multicolumn{2}{r}{10.043}       & \multicolumn{2}{r}{0.01}         & \multicolumn{2}{r}{-0.658}         & \multicolumn{2}{r}{0.055}        \\
\textbf{d$_{yz}$}   & \multicolumn{2}{r}{-0.868}       & \multicolumn{2}{r}{0.01}         & \multicolumn{2}{r}{10.011}       & \multicolumn{2}{r}{-0.76}          & \multicolumn{2}{r}{0.152}        \\
\textbf{d$_{x2y2}$} & \multicolumn{2}{r}{-0.449}       & \multicolumn{2}{r}{-0.658}       & \multicolumn{2}{r}{-0.76}        & \multicolumn{2}{r}{0.024}          & \multicolumn{2}{r}{-0.03}        \\
\textbf{d$_{xy}$}   & \multicolumn{2}{r}{-0.182}       & \multicolumn{2}{r}{0.055}        & \multicolumn{2}{r}{0.152}        & \multicolumn{2}{r}{-0.03}          & \multicolumn{2}{r}{-0.173}      
\end{tabular}
    
    \label{SItable25}
\end{table}

\begin{table}[H]
    \centering
    \caption{Orbital resolved exchange parameters for the spin polarized optimized structure of CoPS$_{3}$.}
    \begin{tabular}{c|rlrlrlrlrl}
\textbf{J$_{2}$ (meV)}    & \multicolumn{2}{r}{\textbf{d$_{z2}$}} & \multicolumn{2}{r}{\textbf{d$_{xz}$}} & \multicolumn{2}{r}{\textbf{d$_{yz}$}} & \multicolumn{2}{r}{\textbf{d$_{x2y2}$}} & \multicolumn{2}{r}{\textbf{d$_{xy}$}} \\ \hline
\textbf{d$_{z2}$}   & \multicolumn{2}{r}{-0.213}       & \multicolumn{2}{r}{0.02}         & \multicolumn{2}{r}{0.018}        & \multicolumn{2}{r}{0.0}            & \multicolumn{2}{r}{-0.113}       \\
\textbf{d$_{xz}$}   & \multicolumn{2}{r}{0.02}         & \multicolumn{2}{r}{0.106}        & \multicolumn{2}{r}{-0.008}       & \multicolumn{2}{r}{-0.014}         & \multicolumn{2}{r}{0.122}        \\
\textbf{d$_{yz}$}   & \multicolumn{2}{r}{0.018}        & \multicolumn{2}{r}{-0.008}       & \multicolumn{2}{r}{0.109}        & \multicolumn{2}{r}{-0.038}         & \multicolumn{2}{r}{0.128}        \\
\textbf{d$_{x2y2}$} & \multicolumn{2}{r}{0.0}          & \multicolumn{2}{r}{-0.014}       & \multicolumn{2}{r}{-0.038}       & \multicolumn{2}{r}{-0.035}         & \multicolumn{2}{r}{0.0}          \\
\textbf{d$_{xy}$}   & \multicolumn{2}{r}{-0.113}       & \multicolumn{2}{r}{0.122}        & \multicolumn{2}{r}{0.128}        & \multicolumn{2}{r}{0.0}            & \multicolumn{2}{r}{-0.199}      
\end{tabular}
    
    \label{SItable26}
\end{table}

\begin{table}[H]
    \centering
    \caption{Orbital resolved exchange parameters for the spin polarized optimized structure of CoPS$_{3}$.}
    \begin{tabular}{c|rlrlrlrlrl}
\textbf{J$_{3}$ (meV)}    & \multicolumn{2}{r}{\textbf{d$_{z2}$}} & \multicolumn{2}{r}{\textbf{d$_{xz}$}} & \multicolumn{2}{r}{\textbf{d$_{yz}$}} & \multicolumn{2}{r}{\textbf{d$_{x2y2}$}} & \multicolumn{2}{r}{\textbf{d$_{xy}$}} \\ \hline
\textbf{d$_{z2}$}   & \multicolumn{2}{r}{0.029}        & \multicolumn{2}{r}{-0.135}       & \multicolumn{2}{r}{-0.136}       & \multicolumn{2}{r}{-0.162}         & \multicolumn{2}{r}{-0.027}       \\
\textbf{d$_{xz}$}   & \multicolumn{2}{r}{-0.135}       & \multicolumn{2}{r}{-0.049}       & \multicolumn{2}{r}{-0.018}       & \multicolumn{2}{r}{-0.2}           & \multicolumn{2}{r}{0.051}        \\
\textbf{d$_{yz}$}   & \multicolumn{2}{r}{-0.136}       & \multicolumn{2}{r}{-0.018}       & \multicolumn{2}{r}{-0.049}       & \multicolumn{2}{r}{-0.215}         & \multicolumn{2}{r}{0.051}        \\
\textbf{d$_{x2y2}$} & \multicolumn{2}{r}{-0.162}       & \multicolumn{2}{r}{-0.2}         & \multicolumn{2}{r}{-0.215}       & \multicolumn{2}{r}{-0.089}         & \multicolumn{2}{r}{-0.007}       \\
\textbf{d$_{xy}$}   & \multicolumn{2}{r}{-0.027}       & \multicolumn{2}{r}{0.051}        & \multicolumn{2}{r}{0.051}        & \multicolumn{2}{r}{-0.007}         & \multicolumn{2}{r}{0.011}       
\end{tabular}
    
    \label{SItable27}
\end{table}

\begin{table}[H]
    \centering
    \caption{Orbital resolved exchange parameters for the spin polarized optimized structure of CoPS$_{3}$.}
    \begin{tabular}{c|rlrlrlrlrl}
\textbf{J$_{4}$ (meV)}    & \multicolumn{2}{r}{\textbf{d$_{z2}$}} & \multicolumn{2}{r}{\textbf{d$_{xz}$}} & \multicolumn{2}{r}{\textbf{d$_{yz}$}} & \multicolumn{2}{r}{\textbf{d$_{x2y2}$}} & \multicolumn{2}{r}{\textbf{d$_{xy}$}} \\ \hline
\textbf{d$_{z2}$}   & \multicolumn{2}{r}{-0.014}       & \multicolumn{2}{r}{0.002}        & \multicolumn{2}{r}{0.005}        & \multicolumn{2}{r}{0.068}          & \multicolumn{2}{r}{-0.048}       \\
\textbf{d$_{xz}$}   & \multicolumn{2}{r}{0.002}        & \multicolumn{2}{r}{-0.019}       & \multicolumn{2}{r}{-0.027}       & \multicolumn{2}{r}{-0.002}         & \multicolumn{2}{r}{0.009}        \\
\textbf{d$_{yz}$}   & \multicolumn{2}{r}{0.005}        & \multicolumn{2}{r}{-0.027}       & \multicolumn{2}{r}{-0.014}       & \multicolumn{2}{r}{-0.013}         & \multicolumn{2}{r}{0.016}        \\
\textbf{d$_{x2y2}$} & \multicolumn{2}{r}{0.068}        & \multicolumn{2}{r}{-0.002}       & \multicolumn{2}{r}{-0.013}       & \multicolumn{2}{r}{-0.071}         & \multicolumn{2}{r}{-0.021}       \\
\textbf{d$_{xy}$}   & \multicolumn{2}{r}{-0.048}       & \multicolumn{2}{r}{0.009}        & \multicolumn{2}{r}{0.016}        & \multicolumn{2}{r}{-0.021}         & \multicolumn{2}{r}{0.002}       
\end{tabular}
    
    \label{SItable28}
\end{table}

\begin{table}[H]
    \centering
    \caption{Orbital resolved exchange parameters for the spin polarized optimized structure of CoPS$_{3}$.}
    \begin{tabular}{c|rlrlrlrlrl}
\textbf{J$_{5}$ (meV)}    & \multicolumn{2}{r}{\textbf{d$_{z2}$}} & \multicolumn{2}{r}{\textbf{d$_{xz}$}} & \multicolumn{2}{r}{\textbf{d$_{yz}$}} & \multicolumn{2}{r}{\textbf{d$_{x2y2}$}} & \multicolumn{2}{r}{\textbf{d$_{xy}$}} \\ \hline
\textbf{d$_{z2}$}   & \multicolumn{2}{r}{-3.726}       & \multicolumn{2}{r}{-0.232}       & \multicolumn{2}{r}{-0.228}       & \multicolumn{2}{r}{-1.283}         & \multicolumn{2}{r}{0.014}        \\
\textbf{d$_{xz}$}   & \multicolumn{2}{r}{-0.232}       & \multicolumn{2}{r}{-0.055}       & \multicolumn{2}{r}{0.0}          & \multicolumn{2}{r}{-0.211}         & \multicolumn{2}{r}{0.002}        \\
\textbf{d$_{yz}$}   & \multicolumn{2}{r}{-0.229}       & \multicolumn{2}{r}{0.0}          & \multicolumn{2}{r}{-0.058}       & \multicolumn{2}{r}{-0.183}         & \multicolumn{2}{r}{-0.004}       \\
\textbf{d$_{x2y2}$} & \multicolumn{2}{r}{-1.283}       & \multicolumn{2}{r}{-0.211}       & \multicolumn{2}{r}{-0.183}       & \multicolumn{2}{r}{-0.06}          & \multicolumn{2}{r}{0.002}        \\
\textbf{d$_{xy}$}   & \multicolumn{2}{r}{0.014}        & \multicolumn{2}{r}{0.002}        & \multicolumn{2}{r}{-0.004}       & \multicolumn{2}{r}{0.002}          & \multicolumn{2}{r}{0.002}       
\end{tabular}
    
    \label{SItable29}
\end{table}

\begin{table}[H]
    \centering
    \caption{Orbital resolved exchange parameters for the spin polarized optimized structure of CoPS$_{3}$.}
    \begin{tabular}{c|rlrlrlrlrl}
\textbf{J$_{6}$ (meV)}    & \multicolumn{2}{r}{\textbf{d$_{z2}$}} & \multicolumn{2}{r}{\textbf{d$_{xz}$}} & \multicolumn{2}{r}{\textbf{d$_{yz}$}} & \multicolumn{2}{r}{\textbf{d$_{x2y2}$}} & \multicolumn{2}{r}{\textbf{d$_{xy}$}} \\ \hline
\textbf{d$_{z2}$}   & \multicolumn{2}{r}{-0.483}       & \multicolumn{2}{r}{0.099}        & \multicolumn{2}{r}{0.087}        & \multicolumn{2}{r}{0.001}          & \multicolumn{2}{r}{0.316}        \\
\textbf{d$_{xz}$}   & \multicolumn{2}{r}{0.099}        & \multicolumn{2}{r}{-0.068}       & \multicolumn{2}{r}{0.0}          & \multicolumn{2}{r}{0.08}           & \multicolumn{2}{r}{0.064}        \\
\textbf{d$_{yz}$}   & \multicolumn{2}{r}{0.087}        & \multicolumn{2}{r}{0.0}          & \multicolumn{2}{r}{-0.066}       & \multicolumn{2}{r}{0.108}          & \multicolumn{2}{r}{0.041}        \\
\textbf{d$_{x2y2}$} & \multicolumn{2}{r}{0.001}        & \multicolumn{2}{r}{0.08}         & \multicolumn{2}{r}{0.108}        & \multicolumn{2}{r}{-3.258}         & \multicolumn{2}{r}{-0.01}        \\
\textbf{d$_{xy}$}   & \multicolumn{2}{r}{0.316}        & \multicolumn{2}{r}{0.064}        & \multicolumn{2}{r}{0.041}        & \multicolumn{2}{r}{-0.01}          & \multicolumn{2}{r}{0.196}       
\end{tabular}
    
    \label{SItable30}
\end{table}

\begin{table}[H]
    \centering
    \caption{Orbital resolved exchange parameters for the experimental structure of NiPS$_{3}$.}
    \begin{tabular}{c|rlrlrlrlrl}
\textbf{J$_{1}$ (meV)}    & \multicolumn{2}{r}{\textbf{d$_{z2}$}} & \multicolumn{2}{r}{\textbf{d$_{xz}$}} & \multicolumn{2}{r}{\textbf{d$_{yz}$}} & \multicolumn{2}{r}{\textbf{d$_{x2y2}$}} & \multicolumn{2}{r}{\textbf{d$_{xy}$}} \\ \hline
\textbf{d$_{z2}$}   & \multicolumn{2}{r}{-0.055}       & \multicolumn{2}{r}{-0.008}       & \multicolumn{2}{r}{-0.012}       & \multicolumn{2}{r}{-0.038}         & \multicolumn{2}{r}{0.002}        \\
\textbf{d$_{xz}$}   & \multicolumn{2}{r}{-0.008}       & \multicolumn{2}{r}{0.014}        & \multicolumn{2}{r}{-0.002}       & \multicolumn{2}{r}{0.011}          & \multicolumn{2}{r}{0.001}        \\
\textbf{d$_{yz}$}   & \multicolumn{2}{r}{-0.012}       & \multicolumn{2}{r}{-0.002}       & \multicolumn{2}{r}{0.012}        & \multicolumn{2}{r}{0.007}          & \multicolumn{2}{r}{0.0}          \\
\textbf{d$_{x2y2}$} & \multicolumn{2}{r}{-0.038}       & \multicolumn{2}{r}{0.011}        & \multicolumn{2}{r}{0.007}        & \multicolumn{2}{r}{-0.312}         & \multicolumn{2}{r}{-0.004}       \\
\textbf{d$_{xy}$}   & \multicolumn{2}{r}{0.003}        & \multicolumn{2}{r}{0.001}        & \multicolumn{2}{r}{0.0}          & \multicolumn{2}{r}{-0.004}         & \multicolumn{2}{r}{-0.002}      
\end{tabular}
    
    \label{SItable31}
\end{table}

\begin{table}[H]
    \centering
    \caption{Orbital resolved exchange parameters for the experimental structure of NiPS$_{3}$.}
    \begin{tabular}{c|rlrlrlrlrl}
\textbf{J$_{2}$ (meV)}    & \multicolumn{2}{r}{\textbf{d$_{z2}$}} & \multicolumn{2}{r}{\textbf{d$_{xz}$}} & \multicolumn{2}{r}{\textbf{d$_{yz}$}} & \multicolumn{2}{r}{\textbf{d$_{x2y2}$}} & \multicolumn{2}{r}{\textbf{d$_{xy}$}} \\ \hline
\textbf{d$_{z2}$}   & \multicolumn{2}{r}{-0.161}       & \multicolumn{2}{r}{-0.012}       & \multicolumn{2}{r}{-0.01}        & \multicolumn{2}{r}{0.0}            & \multicolumn{2}{r}{-0.011}       \\
\textbf{d$_{xz}$}   & \multicolumn{2}{r}{-0.012}       & \multicolumn{2}{r}{-0.006}       & \multicolumn{2}{r}{0.001}        & \multicolumn{2}{r}{0.001}          & \multicolumn{2}{r}{0.005}        \\
\textbf{d$_{yz}$}   & \multicolumn{2}{r}{-0.01}        & \multicolumn{2}{r}{0.001}        & \multicolumn{2}{r}{-0.006}       & \multicolumn{2}{r}{0.001}          & \multicolumn{2}{r}{0.005}        \\
\textbf{d$_{x2y2}$} & \multicolumn{2}{r}{0.0}          & \multicolumn{2}{r}{0.001}        & \multicolumn{2}{r}{0.001}        & \multicolumn{2}{r}{-0.1}           & \multicolumn{2}{r}{-0.008}       \\
\textbf{d$_{xy}$}   & \multicolumn{2}{r}{-0.011}       & \multicolumn{2}{r}{0.005}        & \multicolumn{2}{r}{0.005}        & \multicolumn{2}{r}{-0.008}         & \multicolumn{2}{r}{0.017}       
\end{tabular}
    
    \label{SItable32}
\end{table}

\begin{table}[H]
    \centering
    \caption{Orbital resolved exchange parameters for the experimental structure of NiPS$_{3}$.}
    \begin{tabular}{c|rlrlrlrlrl}
\textbf{J$_{3}$ (meV)}    & \multicolumn{2}{r}{\textbf{d$_{z2}$}} & \multicolumn{2}{r}{\textbf{d$_{xz}$}} & \multicolumn{2}{r}{\textbf{d$_{yz}$}} & \multicolumn{2}{r}{\textbf{d$_{x2y2}$}} & \multicolumn{2}{r}{\textbf{d$_{xy}$}} \\ \hline
\textbf{d$_{z2}$}   & \multicolumn{2}{r}{0.015}        & \multicolumn{2}{r}{-0.0}         & \multicolumn{2}{r}{0.0}          & \multicolumn{2}{r}{0.004}          & \multicolumn{2}{r}{0.001}        \\
\textbf{d$_{xz}$}   & \multicolumn{2}{r}{-0.0}         & \multicolumn{2}{r}{-0.0}         & \multicolumn{2}{r}{-0.001}       & \multicolumn{2}{r}{0.011}          & \multicolumn{2}{r}{-0.0}         \\
\textbf{d$_{yz}$}   & \multicolumn{2}{r}{0.0}          & \multicolumn{2}{r}{-0.001}       & \multicolumn{2}{r}{-0.0}         & \multicolumn{2}{r}{0.012}          & \multicolumn{2}{r}{-0.0}         \\
\textbf{d$_{x2y2}$} & \multicolumn{2}{r}{0.004}        & \multicolumn{2}{r}{0.011}        & \multicolumn{2}{r}{0.012}        & \multicolumn{2}{r}{0.053}          & \multicolumn{2}{r}{0.007}        \\
\textbf{d$_{xy}$}   & \multicolumn{2}{r}{0.001}        & \multicolumn{2}{r}{-0.0}         & \multicolumn{2}{r}{-0.0}         & \multicolumn{2}{r}{0.007}          & \multicolumn{2}{r}{-0.0}        
\end{tabular}
    
    \label{SItable33}
\end{table}

\begin{table}[H]
    \centering
    \caption{Orbital resolved exchange parameters for the experimental structure of NiPS$_{3}$.}
    \begin{tabular}{c|rlrlrlrlrl}
\textbf{J$_{4}$ (meV)}    & \multicolumn{2}{r}{\textbf{d$_{z2}$}} & \multicolumn{2}{r}{\textbf{d$_{xz}$}} & \multicolumn{2}{r}{\textbf{d$_{yz}$}} & \multicolumn{2}{r}{\textbf{d$_{x2y2}$}} & \multicolumn{2}{r}{\textbf{d$_{xy}$}} \\ \hline
\textbf{d$_{z2}$}   & \multicolumn{2}{r}{0.204}        & \multicolumn{2}{r}{-0.001}       & \multicolumn{2}{r}{-0.0}         & \multicolumn{2}{r}{0.024}          & \multicolumn{2}{r}{0.003}        \\
\textbf{d$_{xz}$}   & \multicolumn{2}{r}{-0.001}       & \multicolumn{2}{r}{-0.001}       & \multicolumn{2}{r}{0.0}          & \multicolumn{2}{r}{-0.0}           & \multicolumn{2}{r}{0.0}          \\
\textbf{d$_{yz}$}   & \multicolumn{2}{r}{-0.0}         & \multicolumn{2}{r}{0.0}          & \multicolumn{2}{r}{-0.001}       & \multicolumn{2}{r}{0.0}            & \multicolumn{2}{r}{0.0}          \\
\textbf{d$_{x2y2}$} & \multicolumn{2}{r}{0.024}        & \multicolumn{2}{r}{-0.0}         & \multicolumn{2}{r}{0.0}          & \multicolumn{2}{r}{-0.028}         & \multicolumn{2}{r}{-0.009}       \\
\textbf{d$_{xy}$}   & \multicolumn{2}{r}{0.003}        & \multicolumn{2}{r}{0.0}          & \multicolumn{2}{r}{0.0}          & \multicolumn{2}{r}{-0.009}         & \multicolumn{2}{r}{-0.001}      
\end{tabular}
    
    \label{SItable34}
\end{table}

\begin{table}[H]
    \centering
    \caption{Orbital resolved exchange parameters for the experimental structure of NiPS$_{3}$.}
    \begin{tabular}{c|rlrlrlrlrl}
\textbf{J$_{5}$ (meV)}    & \multicolumn{2}{r}{\textbf{d$_{z2}$}} & \multicolumn{2}{r}{\textbf{d$_{xz}$}} & \multicolumn{2}{r}{\textbf{d$_{yz}$}} & \multicolumn{2}{r}{\textbf{d$_{x2y2}$}} & \multicolumn{2}{r}{\textbf{d$_{xy}$}} \\ \hline
\textbf{d$_{z2}$}   & \multicolumn{2}{r}{-5.191}       & \multicolumn{2}{r}{-0.074}       & \multicolumn{2}{r}{-0.061}       & \multicolumn{2}{r}{-2.334}         & \multicolumn{2}{r}{-0.042}       \\
\textbf{d$_{xz}$}   & \multicolumn{2}{r}{-0.074}       & \multicolumn{2}{r}{-0.006}       & \multicolumn{2}{r}{-0.0}         & \multicolumn{2}{r}{-0.031}         & \multicolumn{2}{r}{0.0}          \\
\textbf{d$_{yz}$}   & \multicolumn{2}{r}{-0.061}       & \multicolumn{2}{r}{-0.0}         & \multicolumn{2}{r}{-0.006}       & \multicolumn{2}{r}{-0.024}         & \multicolumn{2}{r}{-0.0}         \\
\textbf{d$_{x2y2}$} & \multicolumn{2}{r}{-2.334}       & \multicolumn{2}{r}{-0.031}       & \multicolumn{2}{r}{-0.024}       & \multicolumn{2}{r}{-0.38}          & \multicolumn{2}{r}{-0.003}       \\
\textbf{d$_{xy}$}   & \multicolumn{2}{r}{-0.042}       & \multicolumn{2}{r}{0.0}          & \multicolumn{2}{r}{-0.0}         & \multicolumn{2}{r}{-0.003}         & \multicolumn{2}{r}{-0.001}      
\end{tabular}
    
    \label{SItable35}
\end{table}

\begin{table}[H]
    \centering
    \caption{Orbital resolved exchange parameters for the experimental structure of NiPS$_{3}$.}
    \begin{tabular}{c|rlrlrlrlrl}
\textbf{J$_{6}$ (meV)}    & \multicolumn{2}{r}{\textbf{d$_{z2}$}} & \multicolumn{2}{r}{\textbf{d$_{xz}$}} & \multicolumn{2}{r}{\textbf{d$_{yz}$}} & \multicolumn{2}{r}{\textbf{d$_{x2y2}$}} & \multicolumn{2}{r}{\textbf{d$_{xy}$}} \\ \hline
\textbf{d$_{z2}$}   & \multicolumn{2}{r}{-0.058}       & \multicolumn{2}{r}{-0.002}       & \multicolumn{2}{r}{-0.002}       & \multicolumn{2}{r}{0.0}            & \multicolumn{2}{r}{0.028}        \\
\textbf{d$_{xz}$}   & \multicolumn{2}{r}{-0.002}       & \multicolumn{2}{r}{0.0}          & \multicolumn{2}{r}{0.0}          & \multicolumn{2}{r}{-0.003}         & \multicolumn{2}{r}{0.0}          \\
\textbf{d$_{yz}$}   & \multicolumn{2}{r}{-0.002}       & \multicolumn{2}{r}{0.0}          & \multicolumn{2}{r}{0.0}          & \multicolumn{2}{r}{-0.004}         & \multicolumn{2}{r}{0.0}          \\
\textbf{d$_{x2y2}$} & \multicolumn{2}{r}{0.0}          & \multicolumn{2}{r}{-0.003}       & \multicolumn{2}{r}{-0.004}       & \multicolumn{2}{r}{-10.997}        & \multicolumn{2}{r}{-0.092}       \\
\textbf{d$_{xy}$}   & \multicolumn{2}{r}{0.028}        & \multicolumn{2}{r}{0.0}          & \multicolumn{2}{r}{0.0}          & \multicolumn{2}{r}{-0.092}         & \multicolumn{2}{r}{-0.008}      
\end{tabular}
    
    \label{SItable36}
\end{table}

\begin{table}[H]
    \centering
    \caption{Orbital resolved exchange parameters for the spin polarized optimized structure of NiPS$_{3}$.}
    \begin{tabular}{c|rlrlrlrlrl}
\textbf{J$_{1}$ (meV)}    & \multicolumn{2}{r}{\textbf{d$_{z2}$}} & \multicolumn{2}{r}{\textbf{d$_{xz}$}} & \multicolumn{2}{r}{\textbf{d$_{yz}$}} & \multicolumn{2}{r}{\textbf{d$_{x2y2}$}} & \multicolumn{2}{r}{\textbf{d$_{xy}$}} \\ \hline
\textbf{d$_{z2}$}   & \multicolumn{2}{r}{0.099}        & \multicolumn{2}{r}{0.009}        & \multicolumn{2}{r}{0.007}        & \multicolumn{2}{r}{-0.05}          & \multicolumn{2}{r}{0.007}        \\
\textbf{d$_{xz}$}   & \multicolumn{2}{r}{0.009}        & \multicolumn{2}{r}{0.004}        & \multicolumn{2}{r}{-0.002}       & \multicolumn{2}{r}{0.044}          & \multicolumn{2}{r}{-0.0}         \\
\textbf{d$_{yz}$}   & \multicolumn{2}{r}{0.007}        & \multicolumn{2}{r}{-0.002}       & \multicolumn{2}{r}{0.002}        & \multicolumn{2}{r}{0.038}          & \multicolumn{2}{r}{-0.0}         \\
\textbf{d$_{x2y2}$} & \multicolumn{2}{r}{-0.05}        & \multicolumn{2}{r}{0.044}        & \multicolumn{2}{r}{0.038}        & \multicolumn{2}{r}{-0.219}         & \multicolumn{2}{r}{-0.003}       \\
\textbf{d$_{xy}$}   & \multicolumn{2}{r}{0.007}        & \multicolumn{2}{r}{-0.0}         & \multicolumn{2}{r}{-0.0}         & \multicolumn{2}{r}{-0.003}         & \multicolumn{2}{r}{-0.004}      
\end{tabular}
    
    \label{SItable37}
\end{table}

\begin{table}[H]
    \centering
    \caption{Orbital resolved exchange parameters for the spin polarized optimized structure of NiPS$_{3}$.}
    \begin{tabular}{c|rlrlrlrlrl}
\textbf{J$_{2}$ (meV)}    & \multicolumn{2}{r}{\textbf{d$_{z2}$}} & \multicolumn{2}{r}{\textbf{d$_{xz}$}} & \multicolumn{2}{r}{\textbf{d$_{yz}$}} & \multicolumn{2}{r}{\textbf{d$_{x2y2}$}} & \multicolumn{2}{r}{\textbf{d$_{xy}$}} \\ \hline
\textbf{d$_{z2}$}   & \multicolumn{2}{r}{-0.17}        & \multicolumn{2}{r}{-0.009}       & \multicolumn{2}{r}{-0.007}       & \multicolumn{2}{r}{0.001}          & \multicolumn{2}{r}{0.08}         \\
\textbf{d$_{xz}$}   & \multicolumn{2}{r}{-0.009}       & \multicolumn{2}{r}{-0.008}       & \multicolumn{2}{r}{-0.002}       & \multicolumn{2}{r}{0.007}          & \multicolumn{2}{r}{0.005}        \\
\textbf{d$_{yz}$}   & \multicolumn{2}{r}{-0.007}       & \multicolumn{2}{r}{-0.002}       & \multicolumn{2}{r}{-0.009}       & \multicolumn{2}{r}{0.007}          & \multicolumn{2}{r}{0.004}        \\
\textbf{d$_{x2y2}$} & \multicolumn{2}{r}{0.001}        & \multicolumn{2}{r}{0.007}        & \multicolumn{2}{r}{0.007}        & \multicolumn{2}{r}{0.054}          & \multicolumn{2}{r}{-0.004}       \\
\textbf{d$_{xy}$}   & \multicolumn{2}{r}{0.08}         & \multicolumn{2}{r}{0.005}        & \multicolumn{2}{r}{0.004}        & \multicolumn{2}{r}{-0.004}         & \multicolumn{2}{r}{0.002}       
\end{tabular}
    
    \label{SItable38}
\end{table}

\begin{table}[H]
    \centering
    \caption{Orbital resolved exchange parameters for the spin polarized optimized structure of NiPS$_{3}$.}
    %
    
    \label{SItable42}
\end{table}

\begin{figure}[ht]
    \centering
    \includegraphics[width=.5\linewidth]{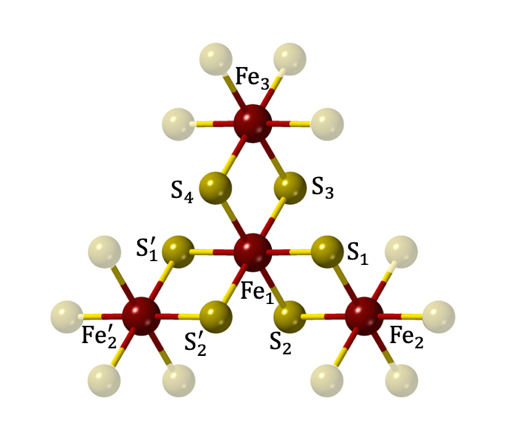}
    \caption{Top view of a cluster of single-layer FePS$_{3}$ centered in Fe$_{1}$ labelling the first Fe atoms neighbours and the S atoms that mediate super-exchange magnetic interactions.}
    \label{FigS4: Sulfur hopping}
\end{figure}

\begin{table}[H]
    \centering
    \caption{Hopping parameters for the crystallographic structure of FePS$_{3}$.}
    %
    
    \label{SItable43}
\end{table}

\begin{table}[H]
    \centering
    \caption{Hopping parameters for the crystallographic structure of FePS$_{3}$.}
    %
    
    \label{SItable44}
\end{table}

\begin{table}[H]
    \centering
    \caption{Hopping parameters for the crystallographic structure of CoPS$_{3}$.}
    %
    
    \label{SItable45}
\end{table}

\begin{table}[H]
    \centering
    \caption{Hopping parameters for the crystallographic structure of CoPS$_{3}$.}
    %
    
    \label{SItable46}
\end{table}

\begin{table}[H]
    \centering
    \caption{Hopping parameters for the crystallographic structure of NiPS$_{3}$.}
    %
    
    \label{SItable47}
\end{table}

\begin{table}[H]
    \centering
    \caption{Hopping parameters for the crystallographic structure of NiPS$_{3}$.}
    %
    \label{SItable}
\end{table}

\newpage
\section{Landau theory of second-order phase transitions and spontaneous magnetostriction}
Magnetostriction is a coupling between magnetic and mechanical parts of our system. This coupling can be described by an energy term in the total free energy of our system \cite{Landau1984}. We can then write the total free energy as 
\begin{equation} \label{SIeq:Landau_Full}
    F-F_0=U_{\rm el}(z)+a(T-T_{\rm N})^{2\beta}L_{i}L_{i}+BL_{i}L_{i}L_{i}L_{i}-\sigma_{ij}(z)\lambda_{ijkl}L_{k}L_{l}.
\end{equation}
Here $F$ is the total free energy of in the AF phase at zero magnetic field, $F_0$ is the free energy of paramagnetic phase, $U_{\text{el}}(z)$ is the elastic energy of a membrane with deflection $z$ at its centre, $T$ is the temperature of our system and $T_{\text{N}}$ is the N\'{e}el temperature, $L_{i}$ are the components of the N\'{e}el vector, $\beta$ is a critical exponent, $a$ and $B$ are positive constants, $\sigma_{ij}(z)$ is the stress tensor and $\lambda_{ijkl}$ is the magnetostriction tensor. The last term couples the stress to the N\'{e}el vector thereby describing the magnetostriction. If we assume the N\'{e}el vector to be fully aligned with the easy axis, Eq. \eqref{SIeq:Landau_Full} simplifies to:
\begin{equation}\label{SIeq:Landau_F}
    F-F_0=U_{\rm el}(z)+a(T-T_{\rm N})^{2\beta}L^2+BL^4-\sigma_{ij}(z)\lambda_{ij}L^2,
\end{equation}
where $L$ is the magnetic order parameter (i.e. the magnitude of the  N\'{e}el vector). For notational convenience we write $\lambda_{ij}$ in dropping the third and fourth index of $\lambda_{ijkl}$ as only the component where $kl$ corresponds to the easy axis contributes.
The elastic energy in a homogeneous membrane is given by \cite{Landau_vol7}
\begin{equation} \label{eq: Elastic Energy General2}
    U_{\rm el} = \int \int \frac{S_{ijkl}}{2} \sigma_{ij}(x,y,z)\sigma_{kl}(x,y,z) dx dy,
\end{equation}
where the integration runs over the in plane dimensions of the membrane, $z$ is the membrane deflection at its centre and should not be confused with the out of plane coordinate. For ease of notation we will not explicitly write the integration and coordinate dependence from here on.
Assuming the membrane thickness does not vary significantly we can take the out of plane component stress component to vanish, $\sigma_{zx} = \sigma_{zy} = \sigma_{zz} = 0$. Eq. \eqref{eq: Elastic Energy General2} then simplifies to
\begin{equation}
         U_{\rm el} = \frac{S_{xxxx}}{2}\sigma_{xx}\sigma_{xx} + \frac{S_{yyyy}}{2}\sigma_{yy}\sigma_{yy} + S_{xxyy}\sigma_{xx}\sigma_{yy} + S_{xxxy}\sigma_{xx}\sigma_{xy} + S_{yyxy}\sigma_{yy}\sigma_{xy} +\frac{S_{xyxy}}{2}\sigma_{xy}\sigma_{xy}.
\end{equation}
Taking our coordinates such that $x,y$ correspond with the principle stress directions all terms containing $\sigma_{xy}$ vanish. This simplifies the elastic energy further to 
\begin{equation} \label{SIeq: noShear General elastic energy}
         U_{\rm el} = \frac{S_{xxxx}}{2}\sigma_{xx}\sigma_{xx} + \frac{S_{yyyy}}{2}\sigma_{yy}\sigma_{yy} + S_{xxyy}\sigma_{xx}\sigma_{yy}.
\end{equation}
Assuming the material has isotropic elastic properties the relevant compliance tensor components are
\begin{equation}
    S_{xxxx}=S_{yyyy}= \frac{1}{E} \text{ and } S_{xxyy}= \frac{-\nu}{E}.
\end{equation}
Substituting this in to Eq. \eqref{SIeq: noShear General elastic energy} we find
\begin{equation}
         U_{\rm el} = \frac{1}{2E}\sigma_{xx}\sigma_{xx} +\frac{1}{2E}\sigma_{yy}\sigma_{yy} - \frac{\nu}{E}\sigma_{xx}\sigma_{yy}.
\end{equation}

By taking the derivative of the free energy with respect to either $z$ or $L$ we find the forces acting on these degrees of freedom, $\phi_{L}$ and $\phi_{z}$ respectively to be given by
\begin{align}
    -\phi_{L} = \frac{d (F-F_{0})}{d L} = 2 a(T-T_{\rm N})^{2\beta}L+4 B L^3-2 \sigma_{ij}(z)\lambda_{ij} L ,\label{eq: Force equation L}\\
    -\phi_{z} = \frac{d (F-F_{0})}{d z} = \left( S_{ijkl} \sigma_{ij}(z) +\lambda_{kl} L^2 \right)  \frac{d\sigma_{kl}(z)}{dz}. \label{eq: Force equation z}
\end{align}

\subsection*{Order parameter and critical exponent}
In order to find an equation that describes the order parameter as a function of temperature, we find a solution for equation \eqref{eq: Force equation L} for the case where $\phi_{L}=0$. Aside from the trivial solution $L = 0$, we find for below the transition the additional solution: 
\begin{equation}
    L^2=-\frac{a}{2B}(T-T_{\rm N})^{2\beta}+\frac{ \sigma_{ij}\lambda_{ij}}{2 B},
\end{equation}
which could be rewritten for $T^*_{\rm N}=T_{\rm N}-(\frac{\sigma_{ij}\lambda_{ij}}{a})^{\frac{1}{2\beta}}$ \cite{Siskins2020} as:
\begin{equation}\label{SIeq:ORDER}
    L^2=\frac{a}{2B}(T^*_{\rm N}-T)^{2\beta}.
\end{equation}
This equation now describes the temperature dependence of the order parameter in a critical region near $T_{\rm N}$ with a corresponding critical exponent $\beta$.

\subsection*{Magnetostrictive strain and resonance frequency}
To assess the magnetostriction contribution to strain and thus the frequency of a rectangular membrane resonator, we need to find stiffness of the membrane from its force-deflection equation. In doing that we analyse equation \eqref{eq: Force equation z}. First, we describe strain equation for the rectangular membrane at its centre as:
\begin{subequations}\label{SIeq:strain_z}
    \begin{align}
    \epsilon_{xx}(z)=\epsilon_{0,x}+\frac{c_1}{2}\frac{z^2}{l^2} \\
    \epsilon_{yy}(z)=\epsilon_{0,y}+\frac{c_1}{2}\frac{z^2}{w^2} 
\end{align}
\end{subequations}
where $c_1$ is a geometrical pre-factor that describes the deflection shape of the fundamental mode of vibration \cite{bunch2008mechanical,SiskinsAs2S32019}. For $w \ll l$ we can neglect the $z$ dependence of $\epsilon_{xx}(z)$. Now, we substitute (\ref{SIeq:strain_z}) to \eqref{eq: Force equation z}, using the relation $\sigma_{ij} = C_{ijkl} \epsilon_{kl}$, we find
\begin{equation}\label{eq: Force z only epsilon_xx z dependent isotropic mat}
    -\phi_{z} = \left( \frac{E}{1-\nu^2} \epsilon_{0,y}+ \frac{\nu E}{1-\nu^{2}} \epsilon_{0,x} -\lambda_{yy} L^2 \right) \frac{c_{1}}{w^2} z 
    + \frac{E}{1-\nu^2} \frac{c_1^{2}}{2}\frac{z^3}{w^4},
\end{equation} 
where we used that $C_{xxxx} = C_{yyyy} = \frac{E}{1-\nu^2}$ and $C_{yyxx} = C_{xxyy} = \frac{\nu E}{1-\nu^{2}}$. Eq. \eqref{eq: Force z only epsilon_xx z dependent isotropic mat} becomes
\begin{equation}
    -\phi_{z} =  k_{1} z -\frac{\lambda_{ij} c_1}{w^2} L^2 z + k_{3} z^3
\end{equation}
where $k_{1}$ is the elastic linear stiffness and $k_{3}$ is the cubic elastic stiffness, given by
\begin{align}
    k_{1} &= \frac{E}{1-\nu^2}( \epsilon_{0,y} + \nu\epsilon_{0,x} )\frac{c_{1}}{w^{2}} \\
    k_{3} &= \frac{E}{1-\nu^2} \frac{c_1^{2}}{2 w^4}.
\end{align}
Assuming small deflections we can neglect the $z^3$ contribution in eq. \eqref{eq: Force z only epsilon_xx z dependent isotropic mat} and find that the linear stiffness is changed with respect to the purely elastic case. If we consider a rectangular cavity with it's long axis is parallel to the crystalline axis $b$ or $a$ respectively we find:
\begin{equation}\label{SIeq:force_z}
        -\phi_{z\ \text{b,a}} =\left(k_1-\frac{c_1}{w^2}\lambda_{\text{a,b}}L^2\right)z \,,
\end{equation}
where $\lambda_{\text{a,b}}$ are the phenomenological magnetostriction coefficients, chosen such that to couple $a$ and $b$ crystalline directions and $L$. That leads to a change in the effective linear stiffness $k_{b,a}$:
\begin{equation}\label{SIeq:Landau_k_ab}
  k_{\text{b,a}}=k_1-\frac{c_1}{w^2}\lambda_{\text{a,b}} L^2,
\end{equation}
which can be used to write down the frequency equations using $f_{\text{a,b}}=\frac{1}{2\pi}\sqrt{\frac{k_{\text{a,b}}}{m}}$,
\begin{subequations}\label{SIeq:freq}
\begin{align}
  f_{\text{b}}&\approx\frac{1}{2\pi}\sqrt{\frac{1}{m}\frac{c_{1}}{w^{2}}\left[\frac{E}{1-\nu^2}( \epsilon_{0,\text{a}} + \nu\epsilon_{0,\text{b}} )-\lambda_{\text{a}} L^2\right]},\\
  f_{\text{a}}&\approx\frac{1}{2\pi}\sqrt{\frac{1}{m}\frac{c_{1}}{w^{2}}\left[\frac{E}{1-\nu^2}( \epsilon_{0,\text{b}} + \nu\epsilon_{0,\text{a}} )-\lambda_{\text{b}} L^2\right]},
\end{align}
\end{subequations}
where $m$ is the mass of the membrane. And define magnetostrictive strain by: 
\begin{align}
  \epsilon_{\rm{ms},\text{a}}&=\frac{c_1}{m w^{2}}\lambda_{\text{a}} L^2,\\
  \epsilon_{\rm{ms},\text{b}}&=\frac{c_1}{m w^{2}}\lambda_{\text{b}} L^2.
\end{align}
Taking a difference of the squares of equations~(\ref{SIeq:freq}) and assuming $\epsilon_{0,\text{a}} = \epsilon_{0,\text{b}}$, we arrive at the final equation:
\begin{align}\label{SIeq:freq_diff_order}
    f^2_{\text{b}}-f^2_{\text{a}}=-\frac{1}{4\pi^2}\frac{c_1}{m w^2}\left[\lambda_\text{b}-\lambda_\text{a}\right]L^2,
\end{align}    
which relates antiferromagnetic order parameter $L$ and measured resonance frequencies of orthogonal resonators aligned to crystalline axes $f_{\text{a,b}}$ in ordered phase. Finally, one can show that by plugging equation~(\ref{SIeq:ORDER}) to (\ref{SIeq:freq_diff_order}):
\begin{equation}\label{SIeq:Landau_k_ab}
  f^2_{\text{b}}-f^2_{\text{a}}\propto (T^*_{\rm N}-T)^{2\beta},
\end{equation}
which could be used to fit experimental data to extract critical exponent $\beta$ near the phase transition temperature $T^*_{\rm N}$.
\section{Derivation of anisotropic resonance frequency}\label{SI Note: Derivation of anisotropic resonance frequency}
\begin{figure}[H]
    \centering
    \includegraphics[width = 0.5\linewidth ]{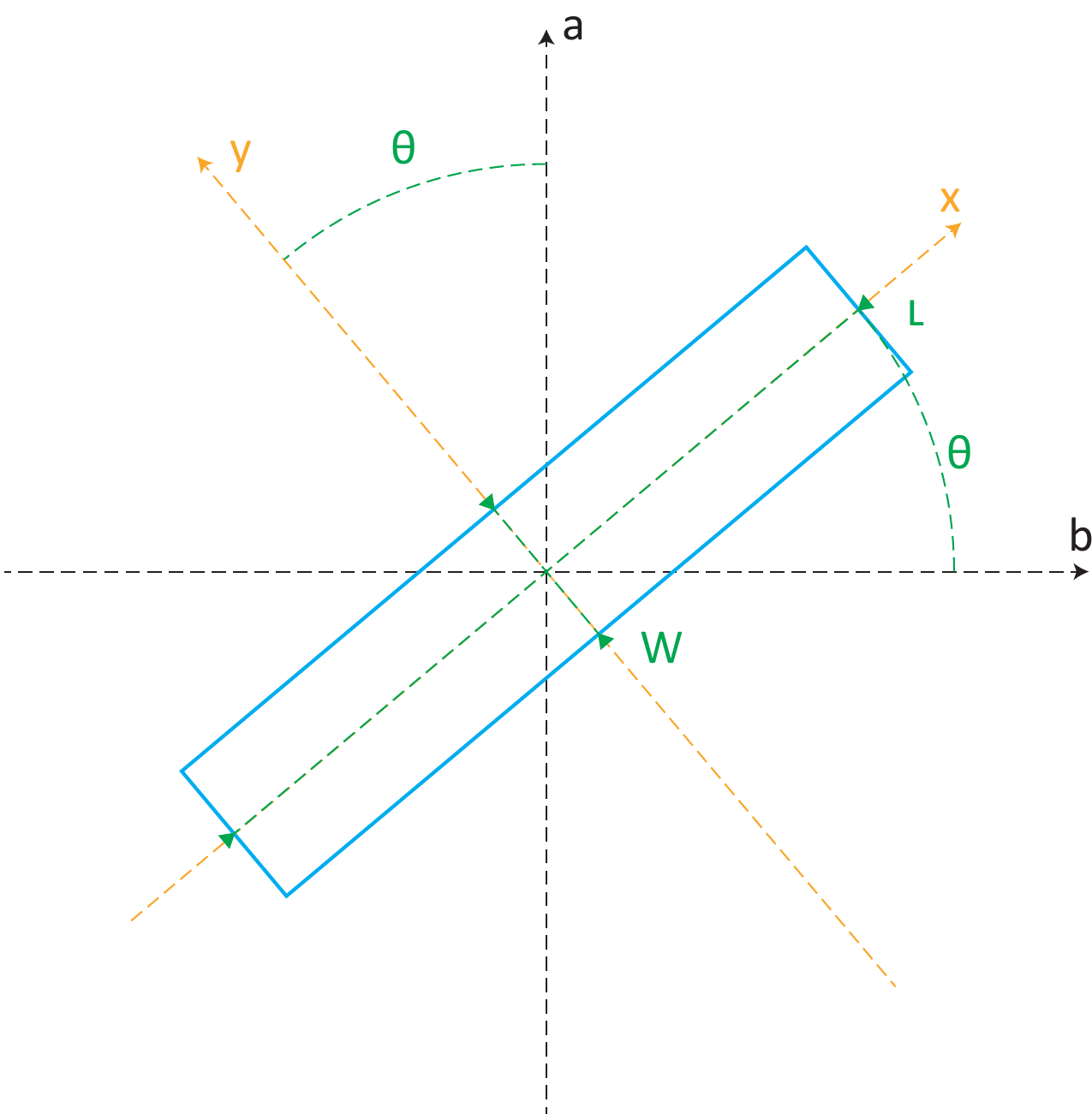}
    \caption{\textbf{Schematic illustration of rectangular membrane} Rectangular membrane of width $w$ and length $l$ oriented with its long side at an angle $\theta$ with respect to the crystalline direction $b$. The $x$-$y$ direction refer to the main directions of the rectangular membrane.}
    \label{fig:SI_rectangle_schematic}
\end{figure}

Here, we derive the general equation of the resonance frequency of a rectangular cavity oriented at an angle $\theta$ with respect to the crystalline axes, as schematically shown in Fig.~\ref{fig:SI_rectangle_schematic}. The global coordinate system is defined by the crystallographic axes \textit{a} and \textit{b}, along which the material deforms resulting in stresses $\sigma_{aa}$ and $\sigma_{bb}$. The longest side of the cavity, with length $l$, can be oriented at an arbitrary angle $\theta$ with respect to \textit{b}.
Let us first consider a cavity oriented parallel to a crystallographic axis. Since the membranes are very thin, we can assume that the stress in the direction perpendicular to the plane is zero, $\sigma_{cc} = 0$, the membrane's stress tensor can be expressed as
\begin{equation}
    \bm{\sigma} = \begin{pmatrix}
        \sigma_{aa} & \sigma_{ab} \\
        \sigma_{ba} & \sigma_{bb}
    \end{pmatrix}_{ab}\,,
\end{equation}
where the subscript $\big( \big)_{ab}$ indicates we expressed the stress tensor in the basis of the crystallographic coordinate system. If we assume that there are no shear forces acting on the crystal lattice, $\sigma_{ab} = \sigma_{ba} = 0$, there will be no shear on cavities oriented along the main crystallographic axes. Now, if we consider a rectangular cavity rotated by $\theta$ with respect to the crystallographic axes, we can define a rotated \textit{xy}-coordinate system oriented along the main axis of the rectangle. To express $\bm{\sigma}$ in this coordinate system we use the tensor transformation rule, $\sigma'_{ij} = q_{ki}q_{lj}\sigma_{kl}$ where $q_{ij}$ are components of the rotation tensor transforming the ab-coordinate system, $\bm{e}$, into the xy-coordinate system, $\bm{e'}$ as $\bm{e}'_{i} = q_{ij} \bm{e}_{j}$. We then get:
\begin{equation}
    \bm{\sigma} = \begin{pmatrix}
        \sigma_{xx} & \sigma_{xy} \\
        \sigma_{yx} & \sigma_{yy}
    \end{pmatrix}_{xy} = \begin{pmatrix}
        \cos^{2}(\theta) \sigma_{aa} + \sin^{2}(\theta) \sigma_{bb} & 
        -\cos(\theta)\sin(\theta) \sigma_{aa} + \sin(\theta)\cos(\theta) \sigma_{bb} \\
        -\sin(\theta)\cos(\theta) \sigma_{aa} + \sin(\theta)\cos(\theta) \sigma_{bb} &
        \sin^{2}(\theta) \sigma_{aa} + \cos^{2}(\theta) \sigma_{bb}
    \end{pmatrix}_{xy}\,,
\end{equation}

Then, the resonance frequency of a rectangular membrane oriented at an angle $\theta$ with respect to the crystallographic axis can be expressed as
\begin{equation}\label{eq:f_theta}
    f_\theta \approx \frac{1}{2}\sqrt{\frac{1}{\rho}\left(\frac{\sigma_{xx}}{l^2} + \frac{\sigma_{yy}}{w^2}\right)}\,.
\end{equation}
In the case of high-aspect ratio membranes ($w\ll l$), Eq. \ref{eq:f_theta} can be approximated to
\begin{equation}\label{eq.resonance_genereal}
    f_\theta \approx \frac{1}{2}\sqrt{\frac{1}{\rho} \frac{\sigma_{yy}}{w^2}} = \frac{1}{2}\sqrt{\frac{1}{\rho w^2} \left(\sin^{2}(\theta) \sigma_{aa} + \cos^{2}(\theta) \sigma_{bb}\right)}\,,
\end{equation}
which is Eq. 3 of the main text.

\begin{figure}
    \centering
    \includegraphics[width =.9\linewidth]{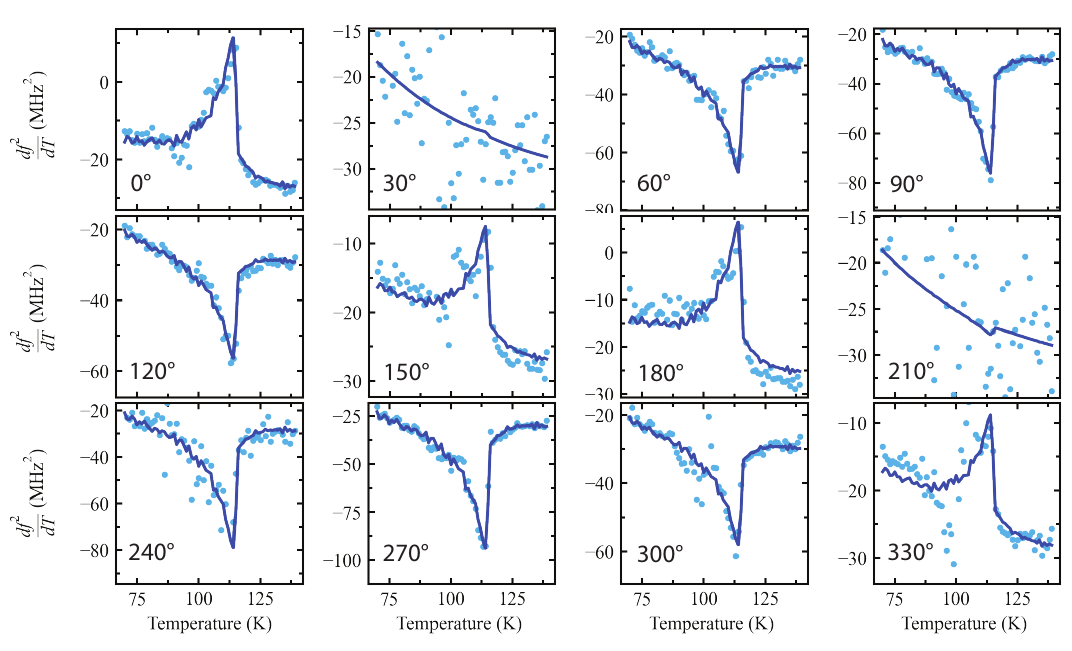}
    \caption{\textbf{Angle-resolved $\dfrac{df^2}{dT}$:} Plot of measured (light blue dots)$\dfrac{df^2}{dT}$  and fit to $b_1 C_\text{Debye} + b_2\dfrac{dL^2}{dT}$ (blue full line) for all angles of a star-shaped array of CoPS$_3$.}
    \label{fig:FigSIdfdT}
\end{figure}

\begin{figure}
    \centering
    \includegraphics[width = .6\linewidth]{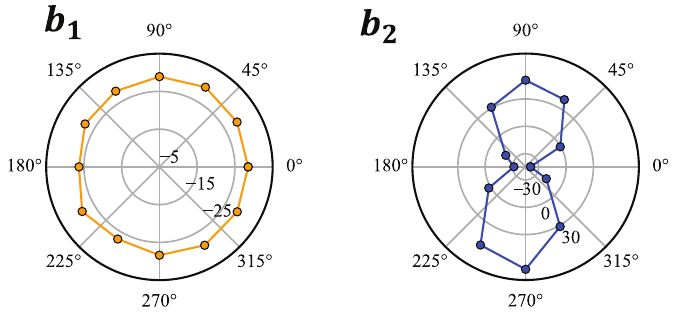}
    \caption{\textbf{Polar plot of fit parameters $b_1$ and $b_2$:} Polar plot of fit parameters $b_1$ and $b_2$ from the fits to $df^2/dT$ shown in Fig. \ref{fig:FigSIdfdT}.}
    \label{fig:my_label}
\end{figure}

Now, let us consider the constitutive equations of the material:
\begin{align}
    c_{1} &= E(\epsilon_{\text{fab},aa} -\epsilon_{\text{th},aa}-\epsilon_{\text{ms},aa})= E \left(\epsilon_{\text{fab},aa} - \int_{T_{0}}^{T_{1}} \alpha_{\text{a}}(T) dT - \lambda_{\text{a}} L^{2}(T_{1})\right) = \sigma_{aa}(T_{1}) - \nu \sigma_{bb}(T_{1}) \\
    c_{2} &= E(\epsilon_{\text{fab},bb} -\epsilon_{\text{th},bb}-\epsilon_{\text{ms},bb})= E \left(\epsilon_{\text{fab},bb} - \int_{T_{0}}^{T_{1}}
    \alpha_{\text{b}}(T) dT - \lambda_{\text{b}} L^{2}(T_{1})\right)
    = \sigma_{bb}(T_{1}) - \nu \sigma_{aa}(T_{1})\,,
\end{align}
where $\epsilon_{\text{fab}}$ is residual fabrication strain at $T=T_{0}$, $\epsilon_{\text{th}}$ and $\epsilon_{\text{ms}}$ are respectively the thermal expansion and magnetostriction contributions to strain, $\alpha$ is the thermal expansion coefficient, $\lambda$ the magnetostriction coefficient and $E$ is the Young's modulus, which is assumed to be isotropic. We can thus write
\begin{align}
    \sigma_{aa} &= c_1 + \nu \sigma_{bb} \\
    \sigma_{bb} &= c_2 + \nu \sigma_{aa}\,,
\end{align}
which can be combined in the following expressions for $\sigma_{aa}$ and $\sigma_{bb}$:
\begin{align}
    \sigma_{aa} &= \frac{c_{1}+\nu c_{2}}{1-\nu^{2}}\\
    \sigma_{bb} &= \frac{c_{2}+\nu c_{1}}{1-\nu^{2}}\,.
\end{align}
We can now rewrite Eq. \ref{eq.resonance_genereal} in terms of the different contributions to strain, i.e. residual strain from fabrication ($\epsilon_{\text{fab}}$), thermal expansion ($\propto \alpha$) and magnetostriction ($\propto \lambda$):
\begin{align}
    f_\theta(T) &= \frac{1}{2}\sqrt{\frac{E}{\rho w^2(1-\nu^2)}\left[\sin^2\theta(c_{1}+\nu c_2)+\cos^2\theta(c_2+\nu c_1)\right]}\\
    & = \frac{1}{2}\sqrt{\frac{E}{\rho w^2(1-\nu^2)}[(\sin^2\theta+\nu\cos^2\theta)(\epsilon_{\text{fab},aa}-\epsilon_{\text{th},aa}-\epsilon_{\text{ms},aa}) + (\cos^2\theta+\nu\sin^2\theta)(\epsilon_{\text{fab},bb}-\epsilon_{\text{th},bb}-\epsilon_{\text{ms},bb})]}\,. \label{eq: f0(theta) arbitrary theta}
\end{align}
Which is consistent with Eq. \eqref{SIeq:freq}. We can eliminate the pretension $\epsilon_{\text{fab}}$ terms by considering $\tilde{f_\theta}(T) = f_\theta(T) - f_\theta(T_{0})$.
In the following, we assume that the only anisotropic temperature-dependent contribution to the total strain comes from magnetostriction, thus we take $\epsilon_{\text{th},aa} = \epsilon_{\text{th},bb} = \epsilon_{\text{th}}$. 
By definition of $\theta$ we have $b\rightarrow\theta = 0^{\circ}$ and $a\rightarrow\theta = 90^{\circ}$. From Eq. \eqref{eq: f0(theta) arbitrary theta}, we then find that $\tilde{f}^2_{\text{a}}-\tilde{f}^2_{\text{b}}$ becomes
\begin{align}
    \tilde{f}^2_{\text{a}}-\tilde{f}^2_{\text{b}} &= \frac{E}{4\rho w^2(1+\nu)} \big(-\epsilon_{\text{ms},aa}+\epsilon_{\text{ms},bb} \big)\\
    &= -\frac{E}{4\rho w^2(1+\nu)}(\lambda_{\text{a}}-\lambda_{\text{b}})L^2 \label{eq:order_parameter_frequency}
\end{align}
from which we can directly extract the order parameter. 
The thermal expansion contribution to strain $\epsilon_\alpha$ is proportional to the integral over the temperature of the thermal expansion coefficient $\alpha$, which is proportional to the Debye specific heat, $C_\text{Debye}$, via the Gr\"{u}nesen parameter. Thus, the derivative with respect to temperature of $f_\theta^2$
\begin{align}
    \frac{df_\theta^2}{dT} &= \frac{E}{4\rho w^2(1-\nu^2)}\left[(\sin^2\theta+\nu\cos^2\theta)\left(-\alpha -\lambda_{\text{a}}\frac{dL^2}{dT}\right) + (\cos^2\theta+\nu\sin^2\theta)\left(-\alpha -\lambda_{\text{b}}\frac{dL^2}{dT}\right)\right]\\
    &= \frac{-E}{4\rho w^2(1-\nu^2)}\left[ \alpha(1+\nu)+ \left(\sin^2\theta(\lambda_{\text{a}} +\nu\lambda_{\text{b}})\frac{dL^2}{dT} +\cos^2\theta(\lambda_{\text{b}}+\nu\lambda_{\text{a}}) \frac{dL^2}{dT}\right)\right]\,.
\end{align}
can be fitted to $b_1 C_\text{Debye} + b_2\dfrac{dL^2}{dT}$ where $b_1$ and $b_2$ are fit parameters, and $\dfrac{dL^2}{dT}$ is estimated from Eq. \ref{eq:order_parameter_frequency}. The results of these fits along to measured data of $\dfrac{df_\theta^2}{dT}$ are shown in Fig. \ref{fig:FigSIdfdT}.Polar plot of the resulting $b_1(\theta)$ and $b_2(\theta)$ are shown in Fig. \ref{fig:my_label}, which confirm that the thermal contribution to strain does not exhibit significant anisotropic behavior.
From Eq. 42 and 40, the expected angle dependance of the parameter $b_2$ is
\begin{equation}
    b_2(\theta) = \left(\frac{E}{4\rho w^2(1-\nu^2)}\right)^2(1-\nu)(\lambda_{\text{a}}-\lambda_{\text{b}})[(\sin^2\theta(\lambda_{\text{a}} +\nu\lambda_{\text{b}}) +\cos^2\theta(\lambda_{\text{b}}+\nu\lambda_{\text{a}})]\,,
\end{equation}
which we use to fit $b_2(\theta)$ in Fig.\ref{fig:my_label} to $A\sin^2\theta + B\cos^2\theta$ where
\begin{equation}
    \frac{A}{B} = \frac{\lambda_{\text{a}} +\nu\lambda_{\text{b}}}{\lambda_{\text{b}} +\nu\lambda_{\text{a}}}\,.
\end{equation}
The fit yields $A/B = -2.062$ and -1.798 for CoPS$_3$ and $A/B = -5.025$ and -8.695 for FePS$_3$.

\section{Anisotropic resonance frequency of F\MakeLowercase{e}PS$_3$ resonators}
\begin{figure*}[ht]
    \centering
    \includegraphics[width = \linewidth]{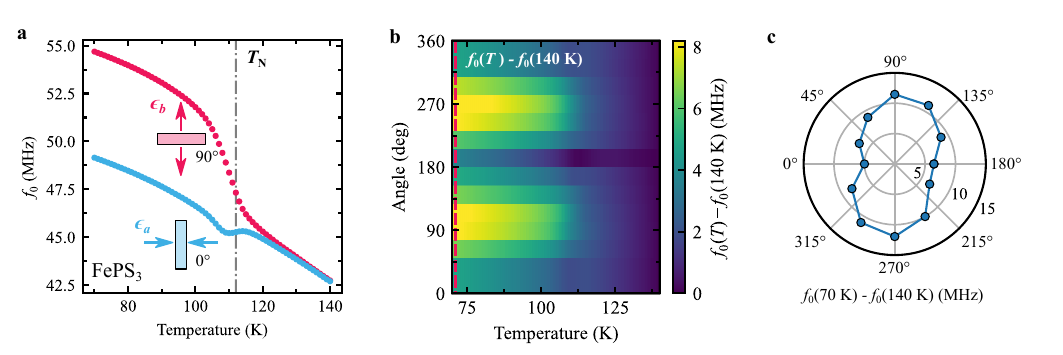}
    \caption{\textbf{Angle-resolved resonance frequency data of FePS$_3$ membranes.} \textbf{a} Temperature dependence of $f_{\text{res}}$ of a FePS$_3$ rectangular membrane orientated at 0$^{\circ}$ (blue) and 90$^{\circ}$ (red) with respect to the \textit{b} crystallographic axis. Note that $f_{0,0^{\circ}}$ and $f_{0,90^{\circ}}$ are proportional to the strain $\epsilon_{\text{a}}$ and $\epsilon_{\text{b}}$ respectively (see Eq. 2). The dashed-dotted grey line indicates the transition temperature $T_\text{N}$. \textbf{b} Resonance frequency difference, $f_{\text{res}}(T)-f_{\text{res}}(140\,\mathrm{K})$, as a function of angle $\theta$ with respect to $b$-axis and temperature. \textbf{c} Polar plot of $f_{\text{res}}(T)-f_{\text{res}}(140\,\mathrm{K})$ taken along the red dashed line in (b).}
    \label{fig:SI1}
\end{figure*}

Figure \ref{fig:SI1} shows resonance frequency data measured on FePS$_3$ star-cavity resonators, as presented in Fig.~2 of the main text for CoPS$_3$ and NiPS$_3$ samples. Figure \ref{fig:SI1}a shows the temperature dependence of the resonance frequency of membranes oriented along the $a$-axis (in red) and $b$-axis (in blue). Similarly to CoPS$_3$, the opposite strain along $a$ and $b$ arising from spontaneous magnetostriction results in opposite behaviour of the resonance frequency near the transition temperature $T_\text{N}$. The resulting anisotropic response is further illustrated in the map plot of resonance frequency as a function of temperature and angles in Fig. \ref{fig:SI1}b and in the polar plot of $f_{\text{res}}(T)-f_{\text{res}}(140\,\mathrm{K})$  in Fig. \ref{fig:SI1}c.

\section{Order parameter related frequency difference $\tilde{f}^2_{\text{b}} - \tilde{f}^2_{\theta}$}
\label{sub sec: Lsquare angle}
\begin{figure*}[ht]
    \centering
    \includegraphics{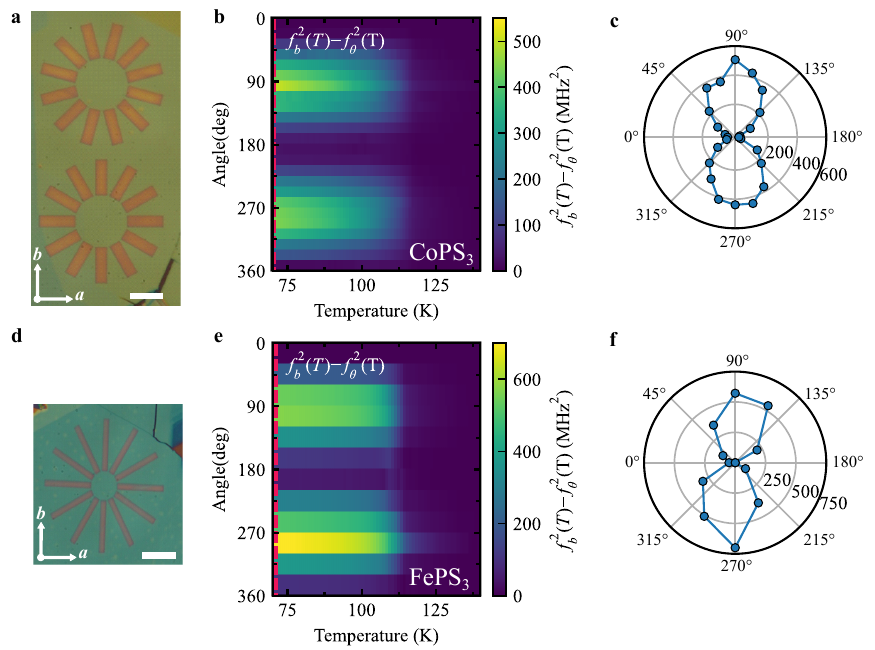}
    \caption{\textbf{Angle-resolved data of $f^2_{\text{b}} - f^2_{\theta} \propto L^2$ of CoPS$_3$ and FePS$_3$ membranes.} \textbf{a,c} Optical image of the CoPS$_3$ (a) and FePS$_3$ (c) resonators. Scale bar 12 $\mu$m. \textbf{b} Resonance frequency difference, $f^2_{\text{b}} - f^2_{\theta} \propto L^2$, as a function of angle $\theta$ and temperature of the CoPS$_3$ sample in (a). \textbf{c} Polar plot of $f^2_{\text{b}} - f^2_{\theta}$ taken along the red dashed line in (b). \textbf{e,f} follows the same structure as (b,c) for the FePS$_3$ sample in (c).}
    \label{fig:SI2}
\end{figure*}

In the main text, we have shown how to relate the difference $f^2_{\text{b}} - f^2_{\text{a}}$ to the antiferromagnetic order parameter through the magnetostriction induced strain at the phase transition $\epsilon_{\text{ms},aa} = \lambda_{\text{a}} L^2$ and $\epsilon_{\text{ms},bb} = \lambda_{\text{b}} L^2$. In general (see derivation in Supplementary Note~2), $\tilde{f}^2_{\text{b}} - \tilde{f}^2_{\theta}$ is also proportional to $L^2$, where $\tilde{f}^2_{\theta}$ is the pretension corrected resonance frequency of a rectangular cavity oriented at an angle $\theta$ with respect to the $b$-axis. We show this quantity for the CoPS$_3$ and FePS$_3$ star-cavity resonators in Fig. \ref{fig:SI2}a,d by plotting $\tilde{f}^2_{\text{b}} - \tilde{f}^2_{\theta}$ as a function of angle and temperature in Fig. \ref{fig:SI2}b,e. Figure \ref{fig:SI2}c,f shows the polar plot of $\tilde{f}^2_{\text{b}} - \tilde{f}^2_{\theta}$ taken along the red dashed line in \ref{fig:SI2}b,e.

This behaviour is observed for the thicker samples ($t > 10$ nm) and it is exploited to have a better estimate of the critical parameters $\beta$ and $T_\text{N}$ as discussed in \ref{sub sec: Lsquare fit}. For thinner resonators every irregularity, like wrinkles or tears, can strongly affect their mode shapes. In some cases, these imperfections can drastically change the resonance frequency of the fundamental mode, as well as its temperature dependence. Therefore, when analysing the critical behaviour of thin flakes, we choose only the most pristine and unaffected membranes out of all fabricated out of a single flake to be sure we are not affected by these irregularities.

\section{Critical curve fit}
\label{sub sec: Lsquare fit}
To extract critical parameters $\beta$ and $T_\text{N}$ shown in Fig. 3 and 4 of the main text, we fit the order parameter related difference $f^2_{\text{b}} - f^2_{\text{a}} \propto L^2$ to the power law $A_\theta(1-T/T_\text{N})^{2\beta}$. The experimental determination of critical parameters is often debated due to the difficulty of extracting from one set of data, three strongly correlated parameters, $\beta$, $T_{\rm N}$ and $A_\theta$. In addition, finite size effects are known to smear the transition which usually results in a non-zero tail of the order parameter in the disordered state and makes it harder to unambiguously determine the critical temperature. Also, the choice of the temperature interval for the fit is not universal and it is often arbitrary.

In order to have a better estimate of the critical exponents from our experiments, we compute $f^2_{\text{b}} - f^2_{\theta}$ for all $\theta$ in a star and fit the data to $A_\theta(1-T/T_\text{N})^{2\beta}$. For each star, we then calculate the average value and standard deviation of the critical parameters $T_{\rm N}$ and $\beta$ weighted by the error from the fit $T_\text{N,err}$ and $\beta_\text{err}$.

\begin{figure}[H]
    \centering
    \includegraphics{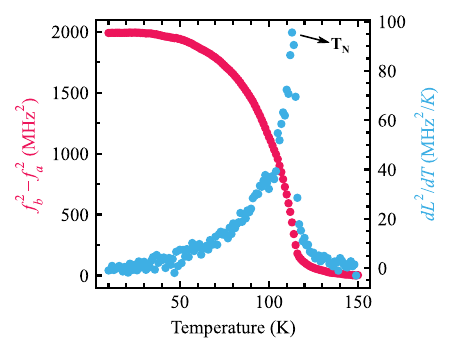}
    \caption{\textbf{First estimate of $T_{\rm N}$ from the derivative of the order parameter.}}
    \label{fig:SI5}
\end{figure}

We start with an initial guess, $T^{*}_\text{N}$, for $T_\text{N}$ by extracting the maximum of the derivative of $f^2_{\text{b}} - f^2_{\theta}$ with respect to temperature as shown in Fig. \ref{fig:SI5}. We then fit the $A_\theta(1-T/T_\text{N})^{2\beta}$ to $f^2_{\text{b}} - f^2_{\theta}$ over the range $[\alpha T^{*}_\text{N},T^{*}_\text{N}]$, for $\alpha$ varying between $[0.85,0.95]$ allowing $A,T_\text{N},\beta$ to vary. We then define the total error for each $\alpha$ to be $T_\text{N,err} + \beta_\text{err}$, where $T_\text{N,err}$, $\beta_\text{err}$ are the standard deviation errors of the fit. We then take the extracted $T_\text{N},\beta$ with corresponding $T_\text{N,err}$, $\beta_\text{err}$ to be ones given the fit corresponding to the $\alpha$ minimizing the total error.
We repeat this process for each $\theta$ yielding a distribution of $T_\text{N},\beta$. We then extract a weighted mean of this distribution as follows:
\begin{equation}
    \overline{\beta} = \frac{1}{N} \sum_{\theta} \beta \frac{\beta_\text{err,min}}{\beta_\text{err}}
\end{equation}
Where $N$ is the number of cavity pairs and $\beta_\text{err,min}$ the $\beta_\text{err}$ of the pairing with smallest $\beta_\text{err}$. We then fit a normal distribution with $\overline{\beta}$ as mean to the distribution of $\beta$ where weigh each $\beta$ by $\frac{\beta_\text{err,min}}{\beta_\text{err}}$, from which we extract the standard deviation. Repeating this process for $T_{N}$. The resulting parameters for each sample are listed in Table \ref{tableS1}.

\renewcommand{\arraystretch}{1.5}
\setlength{\tabcolsep}{15pt}
\begin{table}[H]
\centering
\begin{tabular}{ccccc}
Material & Sample &$t$ (nm)& $\beta$  & $T_\text{N}$ (K) \\ \hline
CoPS$_3$ & 1 & 33  & 0.28 $\pm$ 0.017  & 115.7 $\pm$ 0.38 \\
CoPS$_3$ & 2 & 52  & 0.311 $\pm$ 0.025  & 117 $\pm$ 0.35 \\
CoPS$_3$ & 3 & 52  & 0.298 $\pm$ 0.034  & 116.1 $\pm$ 0.18 \\
CoPS$_3$ & 4 & 8  & 0.195 $\pm$ 0.0446  & 107.8 $\pm$ 2.65 \\
CoPS$_3$ & 5 & 8.6  & 0.218 $\pm$ 0.002  & 102.5 $\pm$  1.3 \\ 
\hline
FePS$_3$ & 1 & 60  & 0.208 $\pm$ 0.0328  & 112.7 $\pm$  0.87 \\ 
FePS$_3$ & 2 & 40  & 0.203 $\pm$ 0.03  & 109.1 $\pm$  0.37 \\
FePS$_3$ & 3 & 10  & 0.194 $\pm$ 0.023  & 110.2 $\pm$  0.48 \\
FePS$_3$ & 4 & 7  & 0.206 $\pm$ 0.047 & 107.9 $\pm$  0.76 \\
\hline
NiPS$_3$ & 1 & 48  & 0.218 $\pm$ 0.016  & 150.7 $\pm$  0.7 \\
\hline
\end{tabular}
\caption{\textbf{Critical exponents of MPS$_3$ samples.} Critical exponents, $\beta$ and $T_\text{N}$, for CoPS$_3$, FePS$_3$ and NiPS$_3$ samples of different thicknesses, extracted following the procedure described in Supplementary Information \ref{sub sec: Lsquare fit}}
\label{tableS1}
\end{table}

\newpage
\putbib